\documentclass[fleqn,usenatbib]{mnras}

\usepackage{newtxtext,newtxmath}

\usepackage[T1]{fontenc}
\usepackage{ae,aecompl}

\usepackage{graphicx}	
\usepackage{amsmath}	
\usepackage{amssymb}	
\usepackage{float}
\usepackage{color}
\usepackage{hyperref}
\usepackage{siunitx}
\usepackage{float}
\usepackage[dvipsnames]{xcolor}

\newcommand{\mtwo}{$M_{\text{200m}}$}
\newcommand{\mstr}{$M_\ast$}

\newcommand{\rtwo}{$r_{\text{200m}}$}
\newcommand{\msun}{M$_{\odot}$}

\newcommand{\lcdm}{$\Lambda$CDM}

\newcommand{\vmax}{$V_\text{max}$}
\newcommand{\vcirc}{$V_\text{circ}$}
\newcommand{\vpeak}{$V_\text{peak}$}

\newcommand{\mq}{\texttt{m11q}}

\newcommand{\mtwos}{$M_{\text{200m}}$ }
\newcommand{\mstrs}{$M_\ast$ }

\newcommand{\rvirs}{$r_{\text{vir}}$ }
\newcommand{\rtwos}{$r_{\text{200m}}$ }
\newcommand{\msuns}{M$_{\odot}$ }

\newcommand{\lcdms}{$\Lambda$CDM }

\newcommand{\vmaxs}{$V_\text{max}$ }
\newcommand{\vcircs}{$V_\text{circ}$ }
\newcommand{\vpeaks}{$V_\text{peak}$ }

\newcommand{\mqs}{\texttt{m11q} }

\newcommand{\rockstars}{\texttt{ROCKSTAR} }

\newcommand{\e}[1]{$\times 10^ {#1}$}

\title[Satellites of LMC Analogs]{Dark and luminous satellites of LMC-mass galaxies in the FIRE simulations}

\author[E.D. Jahn et al.]{
Ethan D. Jahn,$^{1}$\thanks{ethan.jahn@email.ucr.edu}
Laura V. Sales,$^{1}$\thanks{lsales@ucr.edu}
Andrew Wetzel,$^{2}$ 
Michael Boylan-Kolchin,$^3$\newauthor
T.K. Chan,$^4$
Kareem El-Badry,$^5$
Alexandres Lazar,$^6$
James S. Bullock$^6$
\\
$^{1}$Department of Physics and Astronomy, University of California Riverside, 900 University Avenue, Riverside CA 92507, USA\\
$^{2}$Department of Physics, University of California, Davis, CA 95616, USA\\
$^3$Department of Astronomy, The University of Texas at Austin, 2515 Speedway Blvd, Stop C1400, Austin, TX 78712, USA \\
$^4$Department of Physics, Center for Astrophysics and Space Science, University of California at San Diego, 9500 Gilman Dr., La Jolla, CA 92093, USA\\
$^5$Department of Astronomy and Theoretical Astrophysics Center, University of California Berkeley, Berkeley, CA 94720\\
$^6$Department of Physics and Astronomy, University of California, Irvine, CA 92697 USA\\
}

\date{Accepted 2019 August 25. Received 2019 August 22; in original form 2019 July 4.}

\pubyear{2019}

\begin{document}
\label{firstpage}
\pagerange{\pageref{firstpage}--\pageref{lastpage}}
\maketitle

\begin{abstract}
Within \lcdm, dwarf galaxies like the Large Magellanic Cloud (LMC) are expected to host numerous dark matter subhalos, several of which should host faint dwarf companions. Recent Gaia proper motions confirm new members of the LMC-system in addition to the previously known SMC, including two classical dwarf galaxies (\mstrs $ > 10^5$ \msun; Carina and Fornax) as well as several ultra-faint dwarfs (Car2, Car3, Hor1, and Hyd1). We use the Feedback In Realistic Environments (FIRE) simulations to study the dark and luminous (down to ultrafaint masses, \mstrs $\sim$6\e{3} \msun) substructure population of isolated LMC-mass hosts (\mtwo = 1-3\e{11}  \msun) and place the Gaia + DES results in a cosmological context. By comparing number counts of subhalos in simulations with and without baryons, we find that, within 0.2~\rtwo, LMC-mass hosts deplete $\sim$30\% of their substructure, significantly lower than the $\sim$70\% of substructure depleted by Milky Way (MW) mass hosts. For our highest resolution runs ($m_\text{bary}$=880\msun), $\sim 5$-$10$ subhalos form galaxies with \mstrs $ \geq 10^{4}$ \msuns, in agreement with the 7 observationally inferred pre-infall LMC companions. However, we find steeper simulated luminosity functions than observed, hinting at observation incompleteness at the faint end. The predicted DM content for classical satellites in FIRE agrees with observed estimates for Carina and Fornax, supporting the case for an LMC association. We predict that tidal stripping within the LMC potential lowers the inner dark matter density of ultra faint companions of the LMC. Thus, in addition to their orbital consistency, the low densities of dwarfs Car2, Hyd1, and Hyd2 reinforce their likelihood of Magellanic association.
\end{abstract}



\section{Introduction}
The \lcdms structure formation scenario predicts a nested hierarchy of dark matter (DM) halos, subhalos, and sub-subhalos at all mass scales from galaxy clusters to well below the molecular cooling limit of 10$^6$ \msuns \citep{pressSchechter1974,white1978core,tegmark1997, Springel08Aq}. The abundance of substructure is thought to be nearly scale-free and self-similar. That is, the mass function of subhalos takes a universal form when normalized to the mass of the host \citep{gao2004SHpop,kravtsov2004halo,giocoli2008,yang2011}. This means that objects from the most massive halos at the centers of giant clusters to isolated dwarf galaxies should host a similar distribution of DM substructure when normalized properly\footnote{Though, disruption of subhalos and galaxies due to baryons complicates this relation, as described later}. These halos and subhalos act as hosts of galaxy formation, providing potential wells in which gas can accumulate and condense into star forming regions. 
The fact that the relation between stellar mass $M_*$ and halo mass is a near power-law below Milky Way (MW) masses \citep{yang03galform,behroozi10smhm,guo10smhm,moster2013halos,wright2017GAMA}, together with the near invariance of subhalo abundance, means that the number of satellite galaxies 
 normalized to the stellar mass of the central is expected to be independent of host halo mass for  log$_\text{10}$($M_\ast^\text{host}$ / \msun) = 7.25 $ - $ 11.75 \citep{sales2013dwfsats}.

The local group, consisting of the Milky Way (MW), M31, and their numerous satellite galaxies, offers an ideal testing ground for these ideas. For instance, the Large Magellanic Cloud (LMC), the largest satellite of the MW, has long been speculated to host satellites of its own \citep{lyndenbell1995, donghialake08, sales2011MagGal}. Recent observational missions including DES, SMASH, PAN-STARRS, ATLAS, and Gaia have revealed numerous dwarf galaxies in the vicinity of the Magellanic system \citep{bechtol2015DES,drlica2015DES,kimJergen2015hor2,koposov2015DES,torrealba2016Aq2,laevens2015MWsats,helmi2018gaia}, greatly expanding the population of potential LMC-satellites. Based on the orbital properties of these dwarfs, recent works (\citealt{sales2017truMCsats, kallivayalil2018MCsats}, hereafter S17 and K18) have discovered several possible associations of dwarf galaxies to the LMC \citep[see also][]{deason2015LMCsats,jethwa2016magDES,shao2018satacc}.
According to S17 \& K18, there are currently five firm candidates to LMC-system membership: the Small Magellanic Cloud (SMC), Car2, Car3, Hor1, Hyd1, as well as eight promising possible associations awaiting additional proper motion measurements: Dra2, Eri3, Hor2, Hyd2, Phx2, Ret3, Tuc4, and Tuc5. \citet{pardy2019aurigaLMC} have also suggested the possibility of LMC association for the classical dwarf galaxies Carina and Fornax. We expand on this claim in Section \ref{sec:DwarfSatellites}. Each new measurement marks a step closer to a complete picture of the pre-infall satellite population of the LMC, which will greatly inform our theories of galaxy formation.

While it is known that satellite galaxies trace substructure, it is unclear from a theoretical perspective exactly how the population of dark subhalos is mapped to a population of luminous companions. Cosmological N-Body (dark matter only, `DMO') simulations (e.g. \citealt{navarro1996CDMhalos,jenkins1998structure,springel2005millenium,Springel08Aq}) and hydrodynamical baryonic simulations (e.g. \citealt{abadi2003sims,governato2004,sales2007simSats,brooks2013,sawala2016,wetzel2016LATTE}) have been instrumental in making predictions for both central galaxies and smaller-scale subhalo and satellite mass functions, as well as in revealing discrepancies between theoretical predictions of dark matter structure and the observed stellar structure that follows it.

Tension in predictions made by numerical simulations with the observed population of galaxies include the `missing satellites' problem \citep{klypin1999missing, moore1999} and `too big to fail' (\citealt{boylan2011tbtf,tollerud2014tbtf,kirby2014dynamics}; though see also \citealt{Read2006tides,okamotoFrenk2009}), which address this mapping from dark to luminous substructure. Encouragingly, several solutions to such problems have been proposed, mostly invoking a combination of reionization heating, observational incompleteness, and the addition of realistic feedback modeling \citep{bullock2000reionabund,somerville02photo,wetzel2016LATTE,sawala2017,wang2015nihao,gk2018morph,simpson2018auriga}. Furthermore, tides within the host potential affect subhalos. In particular, the addition of the gravitational potential of the central baryonic disk in MW-mass halos has been shown to suppress the presence and survival of dark matter subhalos compared to DMO runs \citep{donghia2010,notsolumpy,graus2018sats}. With a smaller population of subhalos predicted, the number of eligible sites for satellite dwarf galaxies to form is also reduced, highlighting the need for a better understanding of the dark-to-luminous mass mapping predicted within $\Lambda$CDM. Because the central galaxy is the source of this effect, the actual efficacy of this effect depends on the \mstrs - $M_\text{halo}$ ratio. Galaxies near the size of the MW have the highest \mstrs - $M_\text{halo}$ ratio, while galaxies of both higher and lower stellar mass are observed to be hosted by fractionally larger halos, a trend which motivates exploring the magnitude of subhalo depletion by the central galaxy at different host mass scales.

Many of the above studies do not resolve satellite galaxies down to the scale of ultrafaints (UFs; \mstr $\lesssim$ 10$^5$ \msun) - presumably, the most abundant class of galaxies in the universe. Since the satellite population of the LMC can be considered a scaled-down version of that of the MW, it is likely to be dominated by UFs, as the current (incomplete) population of LMC satellites seems to suggest. Due to their shallower gravitational potential, the effects of reionization at this scale of galaxy formation are believed to be stronger than for more massive galaxies like the MW or classical dwarf spheroidals (though feedback effects are weaker since they form proportionally fewer stars). Thus, exploring the ultrafaint population of the LMC (and of LMC-mass hosts in general) is critically important to push the limits of our knowledge of dark-to-luminous substructure mapping into much fainter scales than currently known. 

A possible avenue to overcome numerical resolution limitations in cosmological simulations is to focus on the formation of isolated dwarf galaxies. The lower masses and smaller sizes expected for dwarf halos translate into smaller mass per particle and smaller gravitational softening at a fixed number of particles compared to a more massive halo. For example, \citet{wheeler2015,wheeler2018} were able to study an extremely high resolution population of ultrafaint satellites in simulations of dwarf host halos in the scale \mtwos $ \sim 10^{10}$ \msun. On average, $1-2$ ultrafaints were expected above \mstrs $ \geq 3 \times 10^3$ \msun, making important predictions for the numbers and distribution of ultrafaint dwarfs around dwarf galaxies in the field. Promisingly, larger numbers of ultrafaints shall be expected for more massive dwarf hosts. The inferred virial mass of the LMC, $\sim 10^{11}$ \msuns at the large end of the dwarf galaxy scale, promises to provide numerous substructures with which subhalo abundance can be studied. In addition, its recent infall to the MW ($1-3$ Gyr ago, \citealt{kallivayalil2013}), means its satellite population will remain relatively undisturbed by the tidal field of the MW  \citep{sales2011MagGal, boylankolchin2011dynamics,deason2015LMCsats} and may offer an observational avenue to reconstruct its pre-infall satellite companions.

Previous theoretical works on local group satellite galaxies have predicted that $\sim$30\% of \mstr $\lesssim$ 10$^5$ \msuns satellites of MW-mass hosts fell in as a satellite of a more massive galaxy \citep{wetzelDeasonGK2015}, and that such groups of dwarf galaxies typically disperse in phase space about 5 Gyr after infall to a MW/M31 system \citep{deason2015LMCsats}. In addition, \citet{dooley2017LMCsats} predicts $\sim$ 8 large ultrafaint dwarf satellites of the LMC using reionization and abundance matching models with the Caterpillar simulations \citep{Griffen16caterpillar}. However, such predictions are all based on DMO simulations missing important phenomena expected once the baryons are self-consistently taken into account, for example, SNe feedback, gas hydrodynamics affecting morphology, and tidal effects from the central baryonic disk.

In this work, we present the first analysis of satellite galaxies down to the ultra-faint mass scale in simulations of LMC-mass hosts, using the Feedback In Realistic Environments (FIRE) project. We investigate the luminous and dark substructure of these hosts in hopes of gleaning insight into the real history of the LMC-system. How many satellites, dark and luminous, did the LMC bring with it as it fell into the MW system? What is the mass distribution of these satellites? How have they been shaped by co-evolution with the LMC, and does this differ from satellites/substructures of the MW? Are there ways of constraining membership beyond orbital phase space similarity with the LMC?

Previous works using the FIRE simulations have shown promising results, making them a compelling laboratory to investigate these questions. Recent relevant work includes the observational agreement in classical satellites of MW-mass hosts \citep{wetzel2016LATTE,gk2018morph}, the star formation histories of satellite galaxies in various local-group inspired environemnts \citep{gk2019sfh}, radial profiles of classical satellites \citep{samuel2019radial}, the origin and evolution of the mass-metallicity relation \citep{ma2016metallicity}, chemical abundance distribution of dwarf galaxies \citep{escala2018metaldiffusion}, as well as gas kinematics and morphologies in dwarf galaxies \citep{keb2018angmom,keb2018obskin}.

This paper is organized as follows: In Section \ref{sec:Sims} we descibe the FIRE simulations, incuding the feedback model, the zoom-in technique, halo finding, and  resolution. We also describe our sample of LMC-mass zoom-in runs. In Section \ref{sec:SubSuppression} we describe the effect of the central galaxy population of the abundance of subhalos as compared to DMO runs. In Section \ref{sec:DwarfSatellites} we expand on the analysis of S17, K18, and \citet{pardy2019aurigaLMC} by analyzing Gaia orbital angular momenta of satellites in \citet{helmi2018gaia}. We compare the updated observationally inferred pre-infall mass function of the LMC to the satellite populations in our FIRE runs. We also investigate the structural kinematics of both observed and simulated dwarf satellites, with a focus on ultrafaint dSphs, and consideration of infall time and tidal stripping. We present a summary of this work and concluding remarks in Section \ref{sec:Conclusion}.



\section{Simulations}
\label{sec:Sims}

We analyze a sample of five cosmological zoom-in simulations of LMC-mass host galaxy systems from the Feedback In Realistic Environments project\footnote{\url{http://fire.northwestern.edu}} (FIRE). These runs, implemented in the updated FIRE-2 scheme \citep{hopkins2018fire2}, use the fully conservative cosmological hydrodynamic code \texttt{GIZMO}\footnote{\url{http://www.tapir.caltech.edu/~phopkins/Site/GIZMO.html}} \citep{hopkins2015gizmo}, a multi-method gravity plus hydrodynamics code, in its meshless finite-mass (``MFM'') mode. This is a mesh-free Lagrangian finite-volume Godunov method which automatically provides adaptive spatial resolution while maintaining conservation of mass, energy, and momentum, and excellent shock-capturing and conservation of angular momentum, capturing advantages of both smoothed-particle hydrodynamics (SPH) and Eulerian adaptive mesh refinement (AMR) schemes (for extensive tests, see \citealt{hopkins2015gizmo}). Gravity is solved with an improved version of the Tree-PM solver from GADGET-3 \citep{springel2005gadget2}, with fully-adaptive (and fully-conservative) gravitational force softenings for gas (so hydrodynamic and force softenings are always self-consistently matched), following \citet{price2007}.

\begin{table*}
\centering
 \begin{tabular*}{\textwidth}{l @{\extracolsep{\fill}} c c c c c c c c c c c} 
 Simu- 	&  Resolution  & $M_{\text{200m}}$  & \mstrs  	 & $r_{\text{200m}}$ 	 & $V_{\text{max}}$  & min. \vmax  & $N_{\text{sub}}$ & $N_{\text{sub}}$ & $N_{\text{sub}}$  & $N_\text{lum}$ & Ref. \\ [0.5ex]
 lation	& (\msun) &(\msun) & (\msun) & (kpc) & (km s$^{-1}$) & (km s$^{-1}$) & < 0.2$r_\text{200m}$ & < 0.4$r_\text{200m}$ & < $r_\text{200m}$ & < $r_\text{200m}$\\
 \hline

\texttt{m11c}           &2100 & 1.5e11  & 8.2e8      & 167.4  & 81.0  & 2.5  & 9  & 34   & 118   & 9   &  1  \\

\texttt{m11d}		    &7070 & 2.8e11  & 4.1e9      & 203.9  & 88.8  & 4.0  & 16  & 76  & 257   & 11  &  2  \\
\texttt{m11d-dmo}       &7070 & 3.4e11  &            & 216.5  & 95.4  & 4.8  & 21  & 95  & 333   &     &  3  \\

\texttt{m11e}		    &7070 & 1.5e11  & 1.4e9      & 166.0  & 84.3  & 4.2  & 10  & 31  & 129   & 5   &  2  \\
\texttt{m11e-dmo}       &7070 & 1.8e11  &            & 176.2  & 92.1  & 4.6  & 4   & 34  & 162   &     &  3  \\

\texttt{m11q}		    &880  & 1.5e11  & 3.4e8      & 168.7  & 81.0  & 1.7  & 17  & 56  & 123   & 8   &  4  \\ 
\texttt{m11q-dmo}       &880  & 1.9e11  &            & 182.5  & 91.2  & 1.8  & 20  & 81  & 192   &     &  4  \\

\texttt{m11v}           &7070 & 2.9e11  & 2.4e9      & 210.5  & 84.0  & 4.0  & 11 & 62   & 257   & 6   &  4  \\

\hline 

\texttt{m12b}          &7070 & 1.4e12  & 7.3e10     & 335.1  & 180.9  & 4.2  & 19  & 125  & 544   & 25   &  5   \\
\texttt{m12b-dmo}      &7070 & 1.4e12  &            & 358.8  & 178.8  & 4.7  & 91  & 343  & 1116  &      &  5   \\
\texttt{m12c}          &7070 & 1.4e12  & 5.1e10     & 328.4  & 156.4  & 4.0  & 62  & 285  & 895   & 42   &  5   \\
\texttt{m12c-dmo}      &7070 & 1.3e12  &            & 350.0  & 154.1  & 4.4  & 156 & 537  & 1492  &      &  5   \\
\texttt{m12f}          &7070 & 1.7e12  & 6.9e10     & 354.7  & 183.4  & 4.0  & 34  & 202  & 863   & 31   &  4,5   \\
\texttt{m12f-dmo}      &7070 & 1.8e12  &            & 383.7  & 176.0  & 4.6  & 132 & 519  & 1693  &      &  4,5   \\
\texttt{m12i}          &7070 & 1.2e12  & 5.5e10     & 314.2  & 161.2  & 4.0  & 28  & 194  & 681   & 24   &  4,5   \\
\texttt{m12i-dmo}      &7070 & 1.2e12  &            & 339.3  & 162.3  & 4.8  & 103 & 431  & 1225  &      &  4,5   \\
\texttt{m12m}          &7070 & 1.6e12  & 1.0e11     & 341.6  & 184.3  & 4.3  & 48  & 259  & 900   & 40   &  4,5   \\
\texttt{m12m-dmo}      &7070 & 1.5e12  &            & 364.5  & 171.1  & 4.9  & 116 & 515  & 1622  &      &  4,5   \\
\texttt{m12r}          &7070 & 1.1e12  & 1.5e10     & 304.3  & 136.8  & 4.1  & 53  & 250  & 822   & 28   &  6   \\
\texttt{m12r-dmo}      &7070 & 1.1e12  &            & 322.6  & 146.0  & 4.8  & 95  & 420  & 1192  &      &  6   \\  
\texttt{m12w}          &7070 & 1.1e12  & 4.8e10     & 300.5  & 156.2  & 4.5  & 33  & 181  & 638   & 31   &  6   \\
\texttt{m12w-dmo}      &7070 & 1.2e12  &            & 328.3  & 158.6  & 4.9  & 125 & 448  & 1167  &      &  6   \\
\end{tabular*}
\caption{Properties of the host halo of all simulations analyzed herein for both the N-body dark matter only run (\texttt{-dmo}) and hydrodynamic run (no suffix) at $z=0$. References for additional details on each simulation are shown in the rightmost column. Our primary sample, the LMC-mass \texttt{m11} hosts, are shown in the top block, and our reference sample, the MW-mass \texttt{m12} hosts, are shown below. Resolution refers to the baryonic mass resolution. All quantities are computed from halo catalogs generated by the \texttt{ROCKSTAR} halo finder (calculated using DM particles unless the item explicitly refers to stellar quantities, e.g. \mstr). The minimum \vmaxs is the median value for subhalos of 195<N$_\text{particles}$<205, and serves as an effective resolution limit for substructures in each simulation. This particle number was chosen in accordance with the \citet{hopkins2018fire2} determination of DM convergence radii with the \citet{power03dm} criterion. The virial radius ($r_\text{200m}$) is the radius where the average interior DM density is equal to 200 times the mean matter density in the Universe. The number of subhalos within 0.2$\times$\rtwos, 0.4$\times$\rtwos, and \rtwos of the host are selected as having \vmaxs > 5 km s$^{-1}$, this cutoff value is chosen to include only confidently resolved subhalos in all simulations. Note that \texttt{m11c} and \texttt{m11v} do not have DMO versions. In the last column, we include the number of luminous satellites, defined as \mstrs > 0, though, practically, the stellar mass of the smallest resolved satellite (and hence total number of resolved satellites) varies with resolution. All \mtwos and \mstrs for non-DMO \texttt{m12} simulations were taken from \citet{samuel2019radial}, as they include all mass components in the calculation of \mtwo, as well as only include stellar mass associated with the galactic disk (i.e. excluding stellar mass contained in the stellar halo). References: 1 - \citet{chan2018udgs}; 2 - \citet{keb2018angmom}; 3 - Lazar et al. (in prep); 4 - \citet{hopkins2018fire2}; 5 - \citet{wetzel2016LATTE}; 6 - \citet{samuel2019radial}.}
\label{tab:props}
\end{table*}

FIRE-2 implements a variety of methods for cooling, star formation, and stellar feedback processes. Heating and cooling rates are calculated across 10 - 10$^{10}$K, including \texttt{CLOUDY} ionization states for free-free, photoionization \& recombination, Compton scattering, photoelectric, metal-line, molecular, fine structure, dust collisional, uniform cosmic ray heating, from a spatially uniform, redshift-dependent UV background \citep{faucher2009ionizing}. Local self-shielding is accounted for using a Sobolev approximation. Stars are formed in accordance with \citet{hopkins2013sf}, requiring gas to be locally self-gravitating, self-shielding \citep[following][]{Krumholz2011}, Jeans unstable, and with density $n_H > n_\text{crit} = 1000$ cm$^{-3}$, with all conditions met. Global star formation efficiency is naturally self-regulated by feedback processes, with good observational agreement \citep{orr2018ks}. All newly formed star particles inherit mass and metallicity from their progenitor gas particles.

Stellar feedback quantities are tabulated from the stellar population model STARBUST99 \citep{leitherer1999starburst}, assuming a  \citet{kroupa2001imf} IMF, including supernova Type Ia, II, and stellar winds, as detailed in \citet{hopkins2018fire2} and \citet{hopkins2018sne}. Radiative feedback is modelled with the Locally Extincted Background Radiation in Optically-thin Networks (LEBRON) algorithm \citep{hopkins2018fire2}, accounting for absorbed photon momentum, photo-ionization, and photo-electric heating.

The zoom-in technique \citep{katzwhite1993,onorbe2015fire} is implemented by first simulating a large, low-resolution cosmological box in which a convex Lagrangian region (at initial redshift) is then defined. The region contains all particles within $\sim$5\rvirs at $z=0$ and \textit{not} containing a halo of similar mass to that of the primary, and is then reinitialized with higher resolution. This process is repeated to refine the Lagrangian region until the intended resolution is reached, at which point it is re-simulated with dark matter, gas, and star particles, buffered by a region of lower resolution dark matter particles. Initial conditions\footnote{\url{http://www.tapir.caltech.edu/~phopkins/publicICs/}} are generated with the \texttt{MUSIC} code \citep{hahn2011music}, which implements a second-order Lagrangian perturbation theory to $z=99$.

\begin{figure*}
  \centering 
    \includegraphics[width=\textwidth]{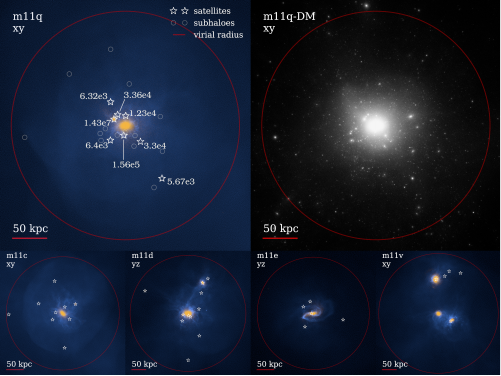}
    
  \caption{Projections of our sample of LMC-analog host galaxies in FIRE, with the virial radius (\rtwo) of each host shown as a red circle. The top left panel shows a detailed projection of the highest resolution run, \mq, with stars in yellow and gas in blue. The fifteen subhalos of highest \vmaxs (down to 9.6 km/s) are shown by grey circles, chosen only to give a visually representative sample of subhalos. All satellite galaxies are shown by grey star markers, with their stellar mass in solar masses (\msun) shown nearby. The dark matter content of the same simulation is shown in the top right panel - a few direct comparisons can be made between bright spots here and subhalo markers on the left. The bottom four panels show the stars and gas of \texttt{m11c}, \texttt{m11d}, \texttt{m11e}, and \texttt{m11v} with satellite galaxies located by star markers. All of our hosts are isolated in the sense that they are not within the realm of influence of a larger halo, such as the Milky Way. However, \texttt{m11v} is an ongoing multimerger, with two neighbors of stellar and virial masses $\lesssim$ that of the host. All other runs are unambiguously isolated. }
  \label{fig:projpanel}
\end{figure*}

Structure and substructure are identified using an updated version of the \rockstars halo finder \citep{behroozi2012rockstar}, which implements 6+1 dimensional phase-space analysis to determine the particles that are gravitationally bound and assign them to (sub)halos. \rockstars assigns mass to halos and subhalos using spherical overdensity calculations relative to a specified threshold, such as the critical density or the average matter density of the universe. Such quantities necessarily become ambiguous and unreliable when quantifying masses of substructure, because the mass density of such embedded objects is by definition above the chosen threshold. To circumvent this ambiguity, we will mostly refer to the maximum circular velocity as an analog for subhalo mass, because it is a more robust and well-defined metric.
For numerical stability, we run \rockstars using only dark-matter particles, and we assign star particles in post-processing using an iterative procedure.
We first select star particles within a DM halo out to 0.8 $r_\text{halo}$ and with velocities less than 2 \vmaxs of the (sub)halo's center of mass velocity, and we iteratively compute stellar position and velocity (making sure the star particles and halo are coincident) until the total mass of star particles, \mstr, converges to <1\%. See \citet{samuel2019radial} for a more detailed description.

\subsection{Resolution Convergence}
\label{ssec:resconv}

All simulations analyzed herein contain no low-resolution particles within $\sim$ 3 \rtwos from the host center. However, the 7070 \msuns resolution runs naturally use fewer particles to represent objects at a given mass than do the higher resolution runs, leading to convergence issues. For this reason, we have determined a minimum value for \vmax, above which subhalos have $\gtrsim$200 DM particles, as listed in Table \ref{tab:props}. We make this cut for all subhalos analyzed herein, and all galaxies naturally fall above this threshold. Using two different resolution runs of \mqs (7070 \msuns \& 880 \msun), we found that the subhalo populations deviate around \vmax $\sim$ 4 km/s, which is below our cutoff in \vmax. This is shown in Figure \ref{fig:convergence}.

\subsection{Our Sample: LMC-like centrals and dark matter cores}
\label{ssec:LMCsample}

Our centrals are selected in the halo mass range \mtwos  = 1.5 $ - $ 3.4\e{11} \msun, where \mtwos refers to the mass measured within \rtwo, defined as the radius at which the mean interior halo density equals 200 times the average matter density of the universe. The stellar masses of the centrals at $z=0$ are in the range \mstrs  = 0.34 $ - $ 4.1\e{9} \msun, compared to the 1.5\e{9} \msuns of the LMC \citep{mcconn2012DG}. These and other properties of our centrals are listed in Table \ref{tab:props}. Our choice of `200m' (as opposed to `200c' which uses the critical density of the universe, $\rho_\text{crit}$) is motivated by its closer physical proximity to the `splashback' radius, in which all (dark) matter that has passed through the core of the halo is enclosed \citep[see][section 2, for a detailed explanation]{wetzel2015accretion}.

We analyze the highest resolution runs available for each system. There are three ``low''-resolution ($m_\text{bary}=$ 7070 \msun) runs: \texttt{m11c, m11d, m11e}; one medium-resolution run ($m_\text{bary}=$ 2100 \msun): \texttt{m11c}; and one high-resolution run ($m_\text{bary}=$ 880 \msun): \texttt{m11q}. For the first part of this analysis, we consider \texttt{m11d}, \texttt{m11e}, and \texttt{m11q} to be our primary focus because they have counterpart DMO simulations, allowing a comparative analysis of subhalo populations between DMO and baryonic versions of the same system. 
We examine all runs in Section \ref{sec:DwarfSatellites}, where we investigate the population of luminous companions around each host. All simulations were run with the core the FIRE-2 hydrodynamics and feedback models, while \texttt{m11q, m11d,} and \texttt{m11e} implemented an additional sub-grid model for the turbulent diffusion of metals in gas. By comparing \texttt{m11q} with and without this model, we have observed no difference in the mass and abundance of satellite galaxies, in agreement with \citet{hopkins2018fire2} and \citet{su2017feedback}, which quantified that including (or not including) metal diffusion physics did not change galaxy-wide properties. Phenomena sensitive to the physics of metal diffusion (such as metallicity gradients) are beyond the scope of this paper, so we proceed considering all runs equivalently. 

To showcase our sample, the top panels in Fig.~\ref{fig:projpanel} show a visualization of the baryons (left) and the dark matter (right) within a region slightly larger than the virial radius of our highest resolution halo \texttt{m11q}. The dark matter component shows a large number of subhalos as expected within \lcdm. On the left, stars (yellow) and gas (blue) have collapsed at the center of this LMC-mass halo to form the central dwarf, with \mstrs $\sim$ 3\e{8} \msun. However, several other satellite subhalos have also formed stars giving rise to the population of dwarf satellites. We highlight them with starred symbols and annotate their corresponding stellar masses. The dark-to-luminous mapping at these low masses is complex, with only a few subhalos forming stars and the rest remaining dark companions. Circles on the left panel indicate the fifteen subhalos of highest \vmaxs that have remained dark. The bottom row shows stars and gas for the rest of our sample and nicely illustrates the variations expected on the luminous companions of the LMC-mass halos.

 Table \ref{tab:props} lists relevant properties of the LMC-analogs analyzed herein. Also listed, for comparison, is the Latte suite of isolated MW-mass hosts \citep[see, e.g., ][hereafter GK17]{wetzel2016LATTE, notsolumpy}. See section \ref{ssec:MWhosts} for more details on these simulations. The selection of all LMC-like hosts was blind with respect to satellite and subhalo population, aside from \texttt{m11v}, which was selected to have an ongoing merger at $z=0$, as can be seen by its two companions, each with \mstr $\gtrsim$ \mstr(SMC) $\sim$ 4\e{8} \msun. All were selected to be isolated from larger halos within $\sim$5 \rtwo. 

\begin{figure*}
\begin{center}
    \centering
    \includegraphics[width=0.98\columnwidth]{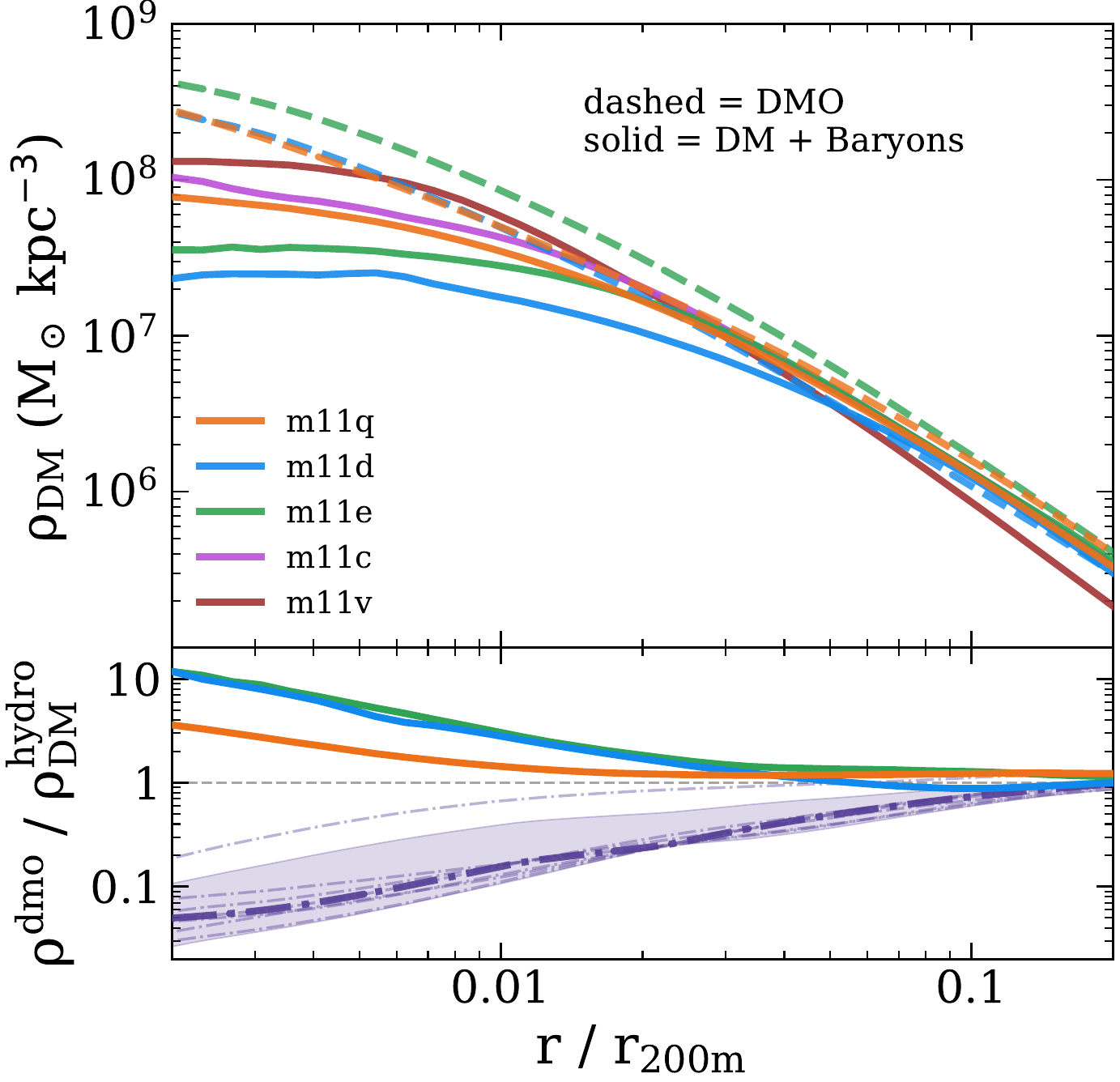}
    \hspace{4mm}
    \includegraphics[width=0.98\columnwidth]{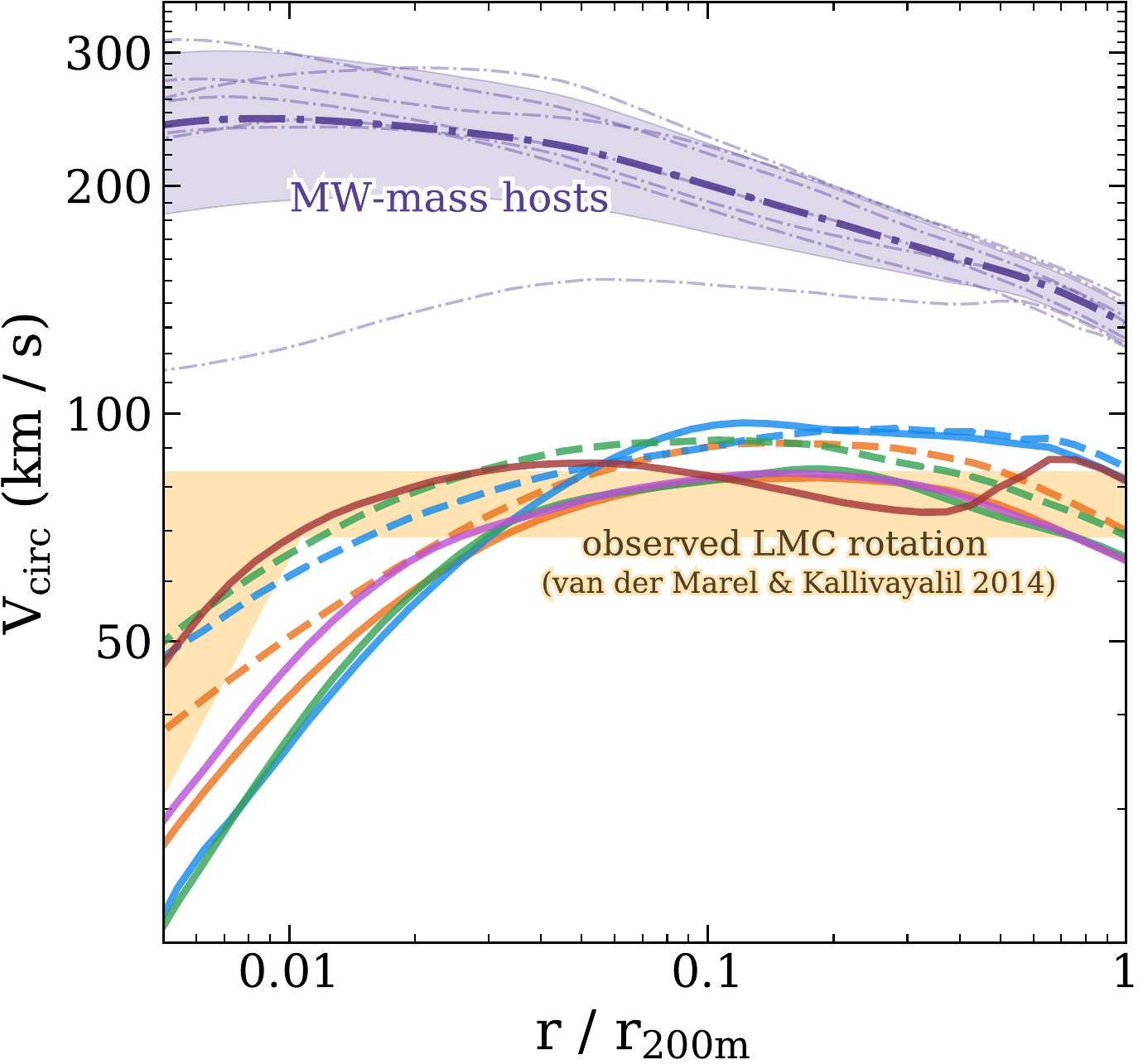}
    \caption{(Left) The top panel shows the DM density of FIRE LMC-mass hosts with distances normalized to \rtwo. Dark Matter Only (DMO) runs are shown as dashed lines while the total mass density for the Hydro runs are shown as solid lines. The bottom panel shows the DMO-to-Hydro density ratio, which quantifies the degree of core-ing. This is shown for MW-mass hosts as the purple dashed-dot line, shaded for the 1$\rm \sigma$ deviation from the mean at $z=0$. All \texttt{m11} runs are heavily cored in comparison to the MW-mass hosts. (Right) Circular velocity profiles of all hosts. \vcircs for baryonic runs is calculated using total mass (DM + stars + gas). The observed circular velocity of the LMC is shaded in orange, representing a $1\rm \sigma$ interval from a parameterized fit of proper motion measurements \citep{vanderMarel2014LMC}. We find good agreement with our simulated LMC-mass hosts for $r$ / \rtwos $\gtrsim0.04$, though all but \texttt{m11v} have somewhat lower circular velocities in the inner regions.}
    \label{fig:density_vcirc}
\end{center}
\end{figure*}

At $z=0$, our LMC-mass centrals all show a cored dark matter density profile (see left panel of Fig.~\ref{fig:density_vcirc}) which contrasts with the denser and cuspy nature of the dark matter only runs (shown for halos \texttt{m11q}, \texttt{m11d} and \texttt{m11e} in dashed lines). This is better indicated in the bottom panel showing the ratio between the dark matter density in the dark matter only (DMO) run to the baryonic run, $\rho_\text{dmo}$ / $\rho_\text{hydro}$,  as a function of radius. Our LMC-mass centrals show $3-10$ times lower densities in the inner regions when the effect of baryons is included, an effect that is much larger than that observed in more massive \mtwos $ \sim 10^{12}$ \msuns hosts run with similar feedback\footnote{\texttt{m11d} and \texttt{m11e} are in the final stages of a merger at $z=0$, making their cores seem larger due to the poorly defined center of each halo.} \citep[dot-dashed purple line showing results from Latte, e.g.][Lazar et al., in prep]{wetzel2016LATTE}.  

The formation of cores in our LMC-like galaxies responds to the rapid removal of self-gravitating gas at the center of the halo due mainly to supernova explosions, which changes temporarily the potential at the center of the halos and leads to the ``heat up'' of the orbits of dark matter particles \citep{navarro1996CDMhalos,penarrubia2010cores,pontGov2012SNe,GK14elvis}. This is consistent with the starbursty nature of star formation reported in FIRE simulations \citep[][]{onorbe2015fire,chan2015FIREcores,keb2016feedback,fitts2017field} and also coincides with the regime where core formation due to the effect of baryons is expected to be maximal \citep{dicintio2014profile,chan2015FIREcores,tollet2016nihaocores}. However, \citet{elbadry2017feedback} showed that the degree of coring is subject to change on relatively short timescales as gas outflows and inflows change. This mass rearrangement also affects the circular velocity profiles, as shown by the right-hand panel of Fig.~\ref{fig:density_vcirc}. Considering the effects of baryons, the maximum circular velocities of our halos are on the high end of agreement with that measured from HST proper motions of stars in the LMC \citep[indicated by the orange shaded region][]{vanderMarel2014LMC}.

\subsection{Our Sample: MW-mass centrals}
\label{ssec:MWhosts}

To highlight the dependence on host mass of many phenomena explored herein, we compare to the Latte suite of seven zoom-in simulations of MW-mass hosts (\texttt{m12b, m12c, m12f, m12i, m12m, m12r,} and \texttt{m12w}; introduced in \citealt{wetzel2016LATTE}) with both baryonic and dark matter only runs, all with a baryonic mass resolution of 7070 \msun. All MW-mass hosts were blindly selected, with the exception of \texttt{m12r} and \texttt{m12w}, which were chosen to have an LMC-mass satellite at $z=0$ in DMO \citep[although, the LMC companion does not necessarily survive to $z=0$ in the baryonic runs, see][]{samuel2019radial}. These simulations confidently resolve subhalos down to \vmaxs $\sim$ 4 km/s, and luminous satellite galaxies down to \mstrs $\sim$ 1\e{5} \msuns (galaxies with \mstrs < 1\e{5} \msuns are mostly found in the higher resolution runs). Halo masses range from (1.1 - 1.7)\e{12} \msun, while stellar masses range from (1.5 - 10)\e{10} \msun. The mean stellar mass to halo mass ratio is $\sim$ 6\e{-2}, compared to $\sim$ 8\e{-3} for the LMC-mass hosts. As shown in Fig.~\ref{fig:density_vcirc}, the central densities of the MW-mass hosts are much greater than those of our LMC-mass hosts, likely due to the effects of baryonic contraction \citep{chan2015FIREcores}. This difference in host density is highly relevant to the discussion on subhalo abundances in the following section.

Any references to the names of various simulations are references specifically to the host/central galaxy of that simulation. The term `companion' refers to a galaxy of nonzero stellar mass within \rtwos of its host, while `subhalo' refers to the self-bound dark matter content of any object - luminous or dark - also within \rtwos of its host. `Satellite' is a collective term for either. As it is a common phrase, we also use the term `virial radius' to refer to \rtwo, and `virial mass' for \mtwo. Note that we do not use the \citet{bryannorman1998} formulation for virial quantities.

\begin{figure*}
	\centering
    \includegraphics[width=\textwidth]{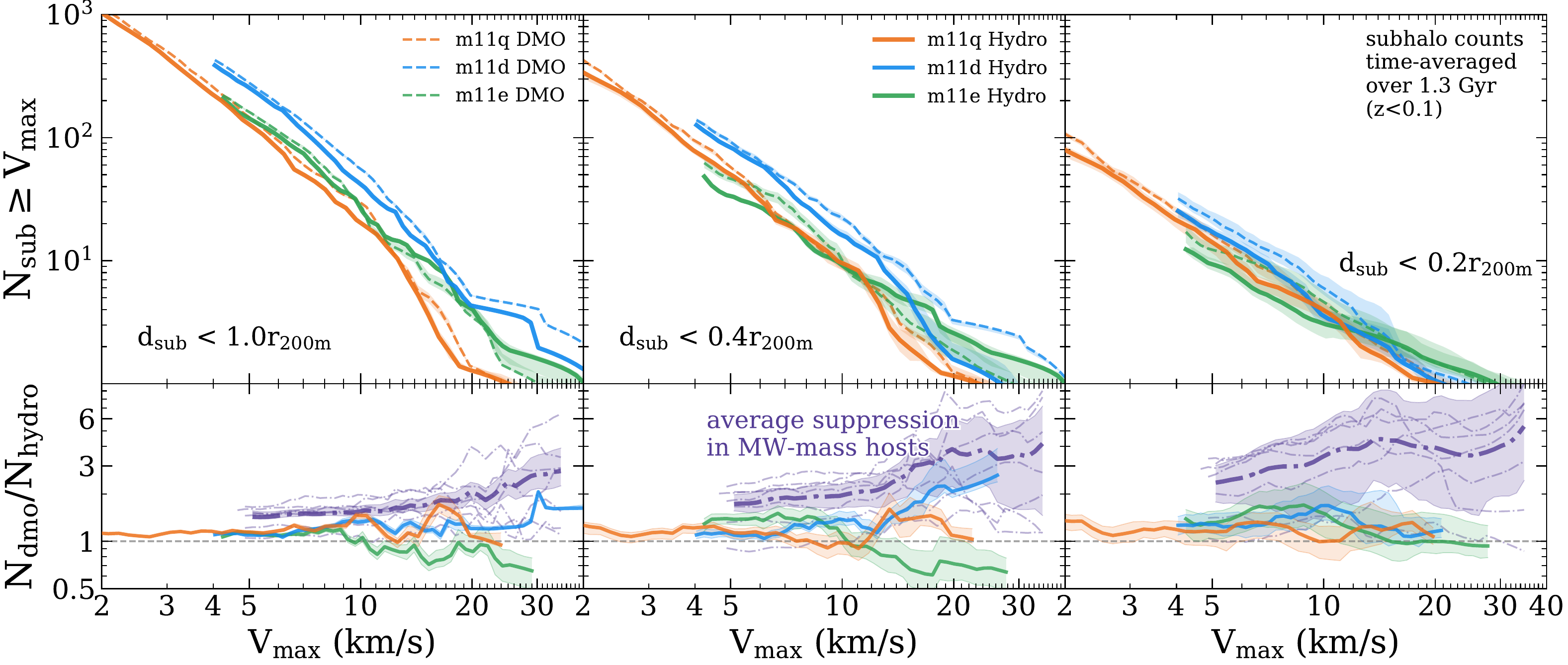}
    \caption{Cumulative time-averaged subhalo count above a given \vmaxs within \rtwos (left), 0.4 \rtwos (center), and 0.2 \rtwos (right) of the host. Counts are averaged over the last $\sim$1.3 Gyr ($z<0.1$) to account for fluctuations in subhalo counts as they pass in and out of each cutoff radius. Each line only extends down to the minimum \vmaxs as listed in Table \ref{tab:props}, and shaded regions indicate 1$\rm \sigma$ deviation from the mean count at a given \vmax. Subhalo \vmaxs in DMO simulations are normalized by $\sqrt{1-f_b}$ to achieve a one-to-one comparison. The bottom panels quantify the magnitude of subhalo depletion as the ratio of the number of subhalos at each \vmaxs in DMO to Hydro. The grey dotted line at $N_\text{dmo}$/$N_\text{hydro} = 1$ represents no depletion (subhalo populations are equivalent in both runs). The purple dashed lines on the bottom panel show the time-averaged depletion of subhalos in MW-mass halos at the corresponding fraction of \rtwos for each host. We observe weaker depletion in the LMC mass halos as compared to the MW mass halos, consistent with a cored potential and lower stellar mass fraction. Table \ref{tab:depletion} shows the average amount of depletion for subhalos of \vmaxs = 10 km/s for the radial cutoffs made here, for both LMC-mass and MW-mass hosts.
    }
    \label{fig:vmax}
\end{figure*}

\section{Suppression of Dark matter subhalos in LMC-like hosts}
\label{sec:SubSuppression}

The number of subhalos above a given \vmax, referred to as the \vmaxs function, is a useful metric to evaluate the abundance and scale of substructure hosted by a central halo. In the last decade, it has been found that the \vmaxs function in MW-mass halos may be significantly suppressed when considering the increased tidal disruption due to the effect of the baryons in the central disk \citep[][GK17]{donghia2010,kelley18phatelvis}. The lower number of surviving subhalos results in a different prediction of the expected number of dwarfs around the MW and in particular for the inner regions where the suppression is maximal. This effect is critical for accurately predicting the radial distribution of satellites around the MW and M31 \citep{samuel2019radial}. These and many other recent studies have primarily been concerned with massive host halos such as the MW. In this section we extend the scope of such inquiry to LMC-mass centrals, with a focus on the suppressing effect of the central baryonic galaxy, by comparing subhalo populations in dark matter only (DMO) simulations to subhalo populations of the same centrals simulated with hydrodynamics \& baryonic physics.

\begin{table}
    \centering
    \begin{tabular*}{\columnwidth}{c @{\extracolsep{\fill}} | c | c | c }
        & $N_\text{dmo}$/$N_\text{hydro}$ &  $N_\text{dmo}$/$N_\text{hydro}$ & $N_\text{dmo}$/$N_\text{hydro}$ \\
        d$_\text{sub}$ <  & \rtwos & 0.4 \rtwos &  0.2 \rtwos  \\
        \hline
         & & & \\
        MW-mass \hspace{2mm} & 1.54$\pm$0.06 & 1.98$^\text{+0.28}_\text{-0.22}$ &  3.51$^\text{+1.82}_\text{-1.72}$ \\
         & & & \\
        LMC-mass \hspace{2mm} & 1.28$^\text{+0.12}_\text{-0.1}$ & 1.14$^\text{+0.35}_\text{-0.23}$ & 1.39$^\text{+0.79}_\text{-0.58}$ \\
         & & & \\
         \hline
    \end{tabular*}
    \caption{Average subhalo depletion ($N_\text{dmo}$/$N_\text{hydro}$) at \vmaxs = 10 km/s  for both MW-mass hosts ($M_\text{halo}\sim$ 10$^{12}$ \msun) and LMC-mass hosts ($M_\text{halo}\sim$ 10$^{11}$ \msun). Each column specifies the cutoff distance inside which subhalos are counted. Among LMC-mass hosts, we observe modest depletion ($N_\text{dmo}$/$N_\text{hydro} \approx 1.1 - 1.4$) of subhalos at all radii, while subhalo depletion is a strong function of distance for MW-mass hosts.}
    \label{tab:depletion}
\end{table}{}

Figure \ref{fig:vmax} shows the time-averaged subhalo \vmaxs function of three LMC-mass hosts (orange - \texttt{m11q}; blue - \texttt{m11d}; green - \texttt{m11e}) as a cumulative count of subhalos at a given \vmaxs and within different radial cuts from \rtwos to 0.2 \rtwos (left to right). Subhalos in each run are plotted from the highest \vmaxs present to the minimum converged \vmax, as listed in Table \ref{tab:props}. Time-averaging was computed over the most recent $\sim$1.3 Gyr, or $z \lesssim 0.1$ by sampling the \vmaxs function of every host at each successive snapshot (66 in total), then computing the mean number of subhalos at each \vmax, including the $\rm 1\sigma$ deviation from the mean (shaded regions). Shading for the average depletion among the Latte suite of MW-mass simulations (bottom panels, purple dashed-dot lines) was calculated by adding the 1$\rm \sigma$ deviations in quadrature, for each run at each \vmax. Within the virial radius of a \mtwos $ \sim 10^{11}$ \msuns halo, on average one can expect order $10$ subhalos with \vmaxs $ \geq 10$ km/s and order $10^2$ above $5$ km/s. We have explicitly checked that the radial distribution of those subhalos follows closely that of the dark matter in the host. 

We additionally include the DMO subhalo populations of all hosts to illustrate how the additional baryonic potential and feedback effects change the distribution and total number of subhalos. This difference is quantified in the bottom panel where the number of subhalos at a given \vmaxs in DMO is divided by that number in Hydro. A number greater than one represents `suppression' or `depletion' of subhalos in the Hydro run, while a number less than one means there are more subhalos at that \vmaxs in Hydro than in DMO, or an `enhancement' of substructure.  

A close inspection of the bottom row in Fig.~\ref{fig:vmax} shows that, on average, all subhalos with \vmaxs < 15 km/s are only slightly suppressed (by a factor of $\sim$ 1.3) in hydro runs compared to the DMO version, an effect that increases only mildly when looking into the inner regions (middle and right panels). While time-averaging is indeed implemented to mitigate the discreteness of the sample, there are not many (<10) satellites with \vmaxs > 15 km/s around LMC-like hosts and Poisson fluctuation may dominate. Take, for example, halo \texttt{m11e} (green). Although the middle panel seems to suggest an increase in the number of subhalos in the hydro run with respect to the DMO, the effect seems localized and it disappears for the inner regions ($r$ < 0.2 \rtwo, right panel). We interpret this as a local fluctuation that results from poor numbers statistics: only 6 subhalos exist with \vmaxs = 10 km/s. We see that the overall trend in our LMC-mass hosts is a mild to negligible suppression of their subhalo populations by their central galaxies.

Also included in the bottom panel is the average suppression of subhalos for all MW-mass Latte hosts within the same fraction of their virial radii as our LMC-mass halos \citep[see also GK17;][]{ samuel2019radial}. In all panels we find that the level of suppression in our LMC-mass halos is significantly smaller than in MW-mass halos. This is most evident in the inner regions $r$ < 0.2 \rtwo, where subhalos are depleted by a factor $\sim 3-6$ in the \texttt{m12}'s compared to a maximum of $\lesssim$ 1.5 in our best resolved (though least cored) halo \texttt{m11q}. These results are mostly independent of the simulation method used. For example, using DMO runs with an added analytic disk \citet{donghia2010} show that MW-mass environments suppress subhalos by a factor $\sim 3$ at $10^7$ \msuns (\vmax$=4-5$ km/s) within 30 kpc of the center, compared to at most a factor $2$ found in our LMC-mass environments. 
The lower impact of baryons on the number of subhalos for LMC-mass hosts is consistent with previous arguments. Using hydrodynamical simulations and comparing to an analytic disk potential, GK17 showed that the primary mechanism for suppression of substructure is the enhanced tidal stripping of subhalos by the gravitational potential of the central galaxy. This is due to two effects in the scale of MW-mass hosts. First, the extra component added to the gravitational forces by the presence of the disk and, second, by the dark matter halo itself becoming more concentrated in the presence of the disk leading to a steeper gravitational potential well. 

We note that we find no correlation between substructure depletion and resolution. For instance, Fig.~\ref{fig:vmax} shows that ``normal'' resolution (\texttt{m11d} and \texttt{m11e}) can show either more or less substructure depletion than the high resolution run \texttt{m11q}. There is therefore not an obvious systematic effect with resolution. Instead, substructure depletion may depend on the specifics of the central baryonic mass and accretion history of each halo. This is consistent with the lack of dependence in the resolution tests for $N_\text{dmo}$/$N_\text{hydro}$ in MW-mass hosts as examined in GK17 and \citet{samuel2019radial}.

In the case of hosts within the \mtwos $ = 10^{11}$ \msuns regime, as analyzed here, the fraction of mass in the disk is much smaller than in MW-mass objects
($\sim$ 1\% average for the LMC-analogs compared to $\sim$ 6\% for the MW-analogs)
partially explaining the lower suppression of substructure. Additionally, the dark matter halos of LMC-hosts in Hydro are predicted to be less dense compared to their DMO counterparts (see Fig.~\ref{fig:density_vcirc}) due to the effect of stellar feedback; opposite to the trend found in MW-mass hosts, which are dominated by baryonic contraction. We conclude that substructure depletion is less significant in dwarfs centrals than expected in $\sim L_*$ galaxies and that hundreds of dark matter subhalos with \vmaxs $ > 5$ km/s are expected to be orbiting around field dwarf galaxies with mass comparable to the LMC.

\subsection{Implications for Observational Subhalo Searches}
\label{ssec:subhalo_obs}

Constraining the degree of subhalo suppression in regions near galaxies has strong implications on the viability of observational subhalo searches that are based on gaps in cold stellar streams. As noted by GK17, if all subhalos within 20 kpc of the MW are suppressed by the baryonic disk, it is less likely for surveys like Palomar-5 and GD-1 \citep{carlberg2012pal5,koposov2010gd1} to detect any interaction with cold dark matter substructure (see also Chapman et al. (in prep), who detected a non-trivial infall rate of subhalos into the inner regions of FIRE MW-mass hosts, but that those subhalos are quickly destroyed). By examining the $z = 0$ radial distribution of substructures with \vmaxs > 10 km/s as seen in Figure \ref{fig:subprof}, we find that, on average, the MW-mass halos host no substructure within 30kpc (though individual halos may occasionally have a few subhalos in the range 20 kpc < d$_\text{sub}$ < 30 kpc). On average, the LMC-mass halos are able to host substructure down to $\sim$20 kpc, while retaining roughly a factor of two more subhalos than MW-mass hosts up to approximately 45kpc. The total number of subhalos hosted by MW-mass halos only surpasses that of LMC-mass halos at $\sim$60-70 kpc. In addition, the overall reduced effect of host-subhalo tidal interactions make LMC-mass hosts somewhat cleaner systems to study substructure than their more strongly interacting MW-mass counterparts.
We therefore argue that the study of cold streams around galaxies with \mstrs $ \sim 10^9$ \msuns may represent a more promising avenue to detect gaps associated to these dark subhalos. 

Of course, the mass range of the substructure is also important. Previous work quantifying the sensitivity of cold stellar structures to disturbances by subhalos (e.g. \citealt{yoon2011cold}) suggest that streams such as Palomar-5 are sensitive to DM substructures in the mass range 10$^7$ to 10$^9$ \msun. The range of median subhalo masses with \vmaxs present in the right panel of Figure \ref{fig:vmax} are 1.1\e{7} to 1.6\e{8} \msun, which, while on the lower end of the quoted sensitivity range, is still within the bounds. This agreement supports our prediction that stellar streams near Magellanic-like systems may have more success that those in the vicinity of the MW-mass objects.

Indirectly, gravitational lensing searches are another avenue to probe dark matter substructure. 
Line ratio anomalies \citep{chiba2002dmsubstructure, metcalfzhou02fluxratios} are strongly influenced by the degree of substructure predicted in the lens system. Novel methods using adaptive optics integral field spectroscopy to measure deviations in quasar narrow line emission \citep{nierenberg2014detect} may require the expected correction due to the baryonic effects. Our results are highly relevant to such searches, suggesting LMC-mass hosts are more likely to maintain a significant amount of substructure in the mass ranges to which such methods are sensitive. 

\begin{figure}
    \includegraphics[width=\columnwidth]{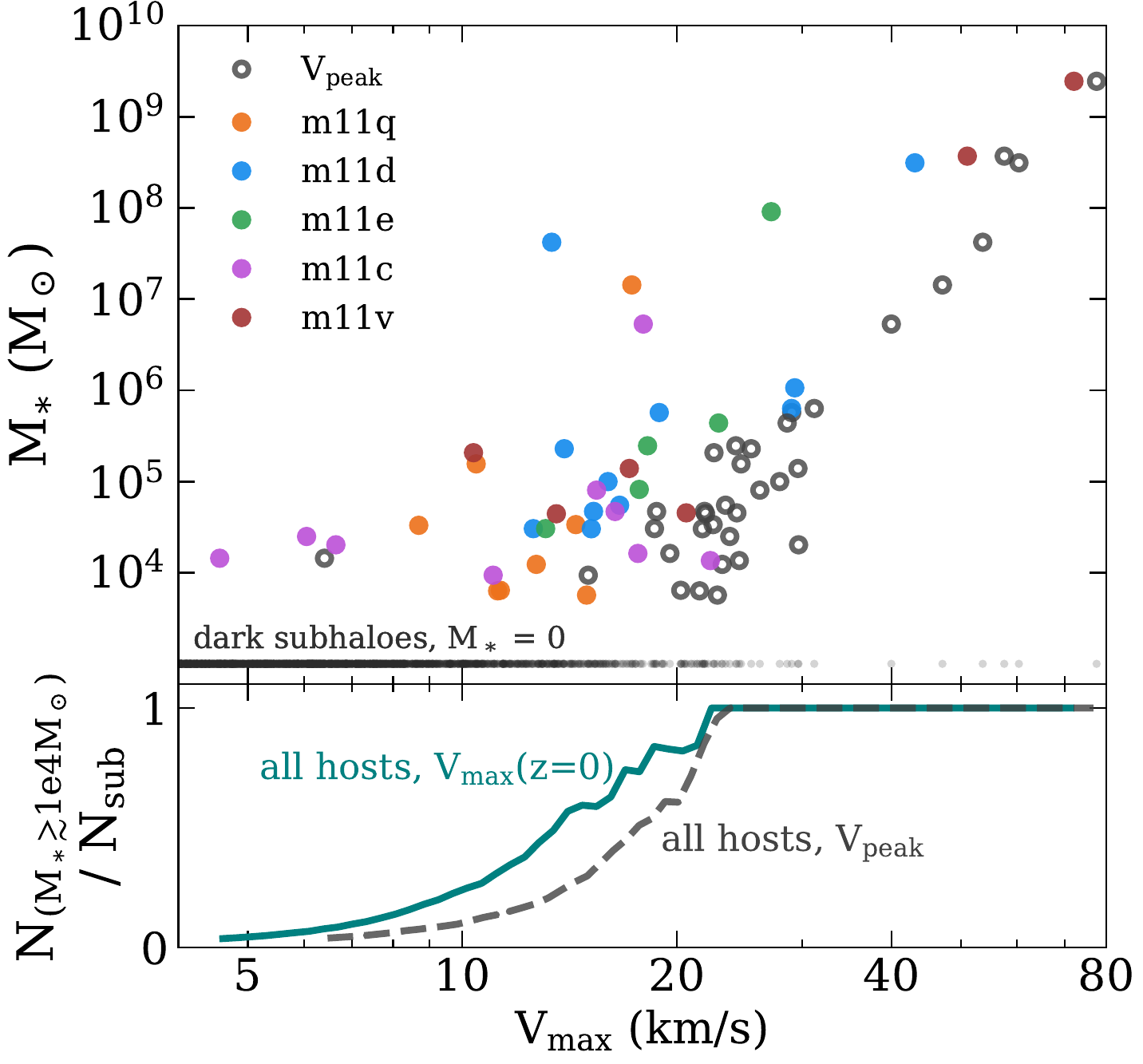}
    \caption{(Top) The $z=0$ \vmaxs (colored) and the peak \vmaxs ever obtained (grey squares) of luminous \mstrs $\gtrsim$ 1\e{4} \msuns companions within the virial radius versus their stellar mass. Subhalos with \vmax$\lesssim$ 20 km/s show a scattered relation with stellar mass, reflecting the stochastic nature of galaxy formation at this scale. (Bottom) The fraction of luminous to dark subhalos at a given \vmaxs or $V_\text{peak}$, using the combined sample of all subhalos with $d_\text{host}$ < \rtwos of all hosts. The galaxy occupation fraction reaches unity for subhalos around \vmaxs $\sim$ 20 km/s, though the distribution of current \vmaxs has been shifted downwards, as expected due to tidal stripping of dark matter.
    }
    \label{fig:vmax_mstar}
\end{figure}

\section{Dwarf satellites in LMC-like hosts}
\label{sec:DwarfSatellites}

A core prediction of \lcdms is that halos and subhalos act as sites of galaxy formation. We therefore expect that some fraction of the surviving dark matter subhalos discussed above in Section \ref{sec:SubSuppression} surrounding our LMC-mass hosts will host a luminous component consisting of stars and gas. The fact that even faint galaxies are much easier to detect than dark subhalos makes the luminous companions of LMC-mass systems a direct and testable prediction of the galaxy formation + $\Lambda$CDM model. In light of recent results by Gaia regarding proper motions of several MW dwarfs, and under the assumption that the availability of 6D information allows one to reconstruct the previous associations of the LMC to other fainter dwarfs \citep{sales2017truMCsats, kallivayalil2018MCsats}, predictions are needed on the number and distribution of visible dwarf satellites expected to orbit around Magellanic dwarfs.

Fig.~\ref{fig:vmax_mstar} highlights the mapping between the stellar mass content and the maximum circular velocities of the subhalos hosting luminous satellites with \mstrs $\gtrsim$ 1\e{4} \msuns of our LMC-mass hosts. At $z=0$, the relation shows significant scatter, meaning that, at a given \vmax, the corresponding stellar content may vary by several orders of magnitude. This effect is partially explained by the tidal stripping of subhalos within the LMC-mass host. This is shown by the gray open symbols indicating the same relation but using the peak value ever obtained for the maximum circular velocity (\vpeak) of each satellite, roughly corresponding to its \vmaxs at infall. A similar increase in scatter in the \mstrs $ - $ \vmaxs relation has been observed for satellites in MW-mass hosts \citep[e.g. ][]{fattahi2013}, however, scatter in the \mstrs $ - $ \vpeaks relation is lower. This could have implications for abundance matching relations at small mass scales \citep[see e.g.][]{gk2017mstrscatter}. We return to tidal effects in sub-section \ref{ssec:dmcontent}. Additional factors contributing to the scatter is the stochasticity of the galaxy formation process near the low-mass end, as well as details of the assembly history of the halo and its inner dark matter density \citep{fitts2017field}.

We note that only a small fraction of the dark matter companions host a luminous dwarf companion (defined here as a galaxy within its host virial radius that has a stellar mass \mstrs $\gtrsim 10^4$ \msuns in \texttt{m11q}). 
The bottom panel of Figure \ref{fig:vmax_mstar} shows the subhalo occupation fraction: the cumulative number of luminous \mstrs $\gtrsim$ 1\e{4} \msuns satellites at a given \vmaxs divided by the cumulative number of dark + luminous subhalos at that \vmax. We have stacked the subhalos and companions of all hosts to achieve a more complete sample across all ranges of \vmax. To account for the different resolution in our runs, we only consider subhalos above the minimum \vmaxs thresholds introduced in Table~\ref{tab:props} for each run.

The occupation fraction for galaxies with \mstrs $\gtrsim$ 1\e{4} \msuns quickly decreases from about unity for \vmaxs $ \geq 20$ km/s to only a few percent at \vmaxs $ \sim 5$ km/s. For reference, our prediction is that about half of the subhalos with \vmaxs$=15$ km/s will host a luminous dwarf, the rest remaining dark or below the \mstrs $=10^4$ \msuns resolved in our runs. These variations are commonly referred to as `stochasticity' in the galaxy formation model, as dark matter subhalos of comparable mass may vary their stellar content by several orders of magnitude, including remaining totally dark, and predicting the exact form of the occupation fraction as a function of \vmaxs must consider many factors of the evolution of subhalos, such as their merger histories and accretion time, as well as external factors such as the onset and end of reionization and the form of the ionizing background.


\begin{table}
\centering
 \begin{tabular*}{\columnwidth}{l @{\extracolsep{\fill}} c c c c} 
Name & Time & $j_x$ & $j_y$ & $j_z$ \\
\hline

LMC          & $t_\text{1p}$	&  -0.97 $\pm$ 0.03	&  0.14 $\pm$ 0.07	& -0.19 $\pm$ 0.10  \\
			 & obs				&  -0.93 $\pm$ 0.06	&  -0.1 $\pm$ 0.03  & -0.36 $\pm$ 0.03	\\
			 
SMC          & $t_\text{1p}$	&  -0.92 $\pm$ 0.05	&  0.04 $\pm$ 0.10  & 0.35 $\pm$ 0.08	\\
			 & obs				&  -0.87 $\pm$ 0.06 & -0.4 $\pm$ 0.04 & -0.27 $\pm$ 0.04 \\
			 
Carina       & $t_\text{1p}$	&  -0.93 $\pm$ 0.12	&  0.25 $\pm$ 0.07  & -0.04 $\pm$ 0.20	\\
			 & obs				&  -0.97$^{+0.13}_{-0.14}$ &  0.17 $\pm$ 0.04  & -0.19 $\pm$ 0.1 \\
		
Fornax       & $t_\text{1p}$ 	&  -0.92 $\pm$ 0.20	&  0.19 $\pm$ 0.11	& 0.20 $\pm$ 0.07 \\
			 & obs				&  -0.96 $^{+0.23}_{-0.22}$ & -0.17$^{+0.13}_{-0.14}$  & 0.24 $\pm$ 0.07  \\
			
Sculptor     & $t_\text{2p}$ 	&  -0.94 $\pm$0.06	&  -0.00 $\pm$ 0.41	& 0.04 $\pm$ 0.05	\\
			 & obs				&  0.99 $\pm$ 0.01  &  -0.03 $^{+0.08}_{-0.07}$  & 0.13 $\pm$ 0.01	\\
			
Sagittarius  & $t_\text{1p}$	& 	-		& -			& -		\\
			 & obs				&  0.05$\pm$ 0.02 	&  -0.99$^{+0.09}_{-0.08}$ & -0.13$^{+0.08}_{-0.09}$ \\
		
Ursa Minor   & $t_\text{1p}$	& 	-		& -			& -		\\
			 & obs				&  -1.0 $\pm$ 0.08 &  -0.09 $\pm$ 0.07 & -0.02 $\pm$ 0.06  \\
		
Leo I        & $t_\text{1p}$	& 	-		& -			& -		\\
			 & obs				&  -0.5 $^{+0.4}_{-0.37} $	&  -0.59$^{+0.5}_{-0.48}$	& -0.64$^{+0.36}_{-0.34}$  \\
		
Sextans      & $t_\text{1p}$	& 	-		& -			& -		\\
			 & obs				&  -0.39 $\pm$ 0.05	&  -0.58 $\pm$ 0.06 & -0.71 $\pm$ 0.06  \\
		
Leo II       & $t_\text{2p}$	&  -0.92 $\pm$ 0.05	& 0.21 $\pm$ 0.21 	& -0.28 $\pm$ 0.15		\\
	 		 & obs				&  -0.13$^{+2.26}_{-2.05}$	& 0.97$^{+2.24}_{-2.12}$ & 0.22 $^{+0.88}_{-0.81}$  \\
	 		
Bootes I     & $t_\text{1p}$	& 	-		& -			& -		\\
			 & obs				&  0.63$^{+0.16}_{-0.15}$ 	&  0.71 $\pm$ 0.11	& -0.31 $\pm$ 0.08 \\
		
Draco        & $t_\text{1p}$	& 	-		& -			& -		\\
			 & obs				&  0.89$^{+0.1}_{-0.09}$ &  0.36 $\pm$ 0.06 & -0.27 $^{+0.08}_{-0.09}$ \\

\end{tabular*}
\caption{Normalized cartesian components of satellite galaxy orbital angular momenta, as predicted by S17 for the first pericentric passage of the LMC (labelled `$t_\text{1p}$') and observed angular momenta as calculated form GAIA proper motions, originally tabulated in \citet{helmi2018gaia} Table C.4 (labelled `obs'). These new measurements show Carina and Fornax as being consistent with a co-infall scenario with the LMC using criteria for Magellanic Cloud system membership as 
defined in S17: $j_x < 0$ and $|j_x| > |j_y|, |j_z|$.}
\label{jtable}
\end{table}

\subsection{New proper motions from Gaia}
\label{ssec:gaia}

From the observational side, determining the dwarf satellites associated to the LMC prior to infall into the MW is not straightforward. Since tidal stripping due to the MW potential has already begun, the material once associated to the LMC in the past does not necessarily cluster around it today. However, because the LMC is inferred to be most likely in its first pericenter passage \citep{kallivayalil2013}, cosmological simulations suggest that the stripped material may still retain its phase-space coherence, opening an avenue to disentangle previous associations \citep{sales2011MagGal, sales2017truMCsats, deason2015LMCsats, jethwa2016magDES}. This coherence means that all subhalos within the LMC at infall are therefore expected to be distributed on the sky following the projection of positions and velocities of the LMC's orbit. Note that final membership requires of a combination of position on the sky, galactocentric distance, and 3D velocity to be satisfied simultaneously.  

This imposes strong constraints on the orbital poles expected for early companions of the LMC and can be used to single out possible associations. This criteria was used in \citet{sales2011MagGal} to conclude that, with the exception of the SMC, no other classical dwarf with available proper motions at that time was consistent with an LMC association. With the arrival of new Gaia DR2 data, proper motions are now available for many dwarfs in the MW, including classical and ultrafaint dwarfs (we assume the cutoff for ultrafaints to be a stellar mass of \mstrs $ < 10^5$ \msun, as in \citealt{bullock2017lcdm}). With the new 6D information, \citealt{kallivayalil2018MCsats} confirmed a likely association for 4 ultrafaints: Car2, Car3, Hor1 and Hyd1. The authors suggest follow up measurements for the confirmation of Dra2 and Hyd2, which have incomplete proper motion information. On the other hand, several of the classical dwarfs have updated proper motion measurements and their membership needs to be re-evaluated. 

We have repeated the analysis in \citet[S11]{sales2011MagGal} but now using the Gaia DR2 proper motions presented in \citep[][H18]{helmi2018gaia}. In particular, for classical dwarfs satisfying the galactocentric distance, position on the sky, and radial velocity constraints, S11 lists predictions for the orbital angular momentum expected in case of association. Following S11, we use a Cartesian coordinate system centered on the Milky Way, with $x$ in the sun-galactic center towards $l = 0$\textdegree, $y$ towards $l = 90$\textdegree in the direction of Galactic rotation, and $z$ coincident with the disk angular momentum, towards $b = 90$\textdegree. For completeness, we list in Table \ref{jtable} the $j_x$, $j_y$ and $j_z$ (all normalized to | $\vec{j}$ |) of all dwarfs included in H18 that were not part of the \citet{kallivayalil2018MCsats} analysis. Errors are propagated from the quoted errors in H18 based on our calculation of $j_x$, $j_y$ and $j_z$. These calculated values are labelled `obs' for each individual galaxy. However, of most relevance to this work are the dwarfs for which S11 had pre-determined possible association based on sky positions and radial velocity. For those cases, we also list the predicted angular momenta for the first pericentric passage ($t_\text{1p}$). Note that Sculptor and Leo II become a possible association only on a second-pericentric passage according to S11 ($t_\text{2p}$). 

\begin{figure*}
    \centering
    
    \includegraphics[width=0.98\columnwidth]{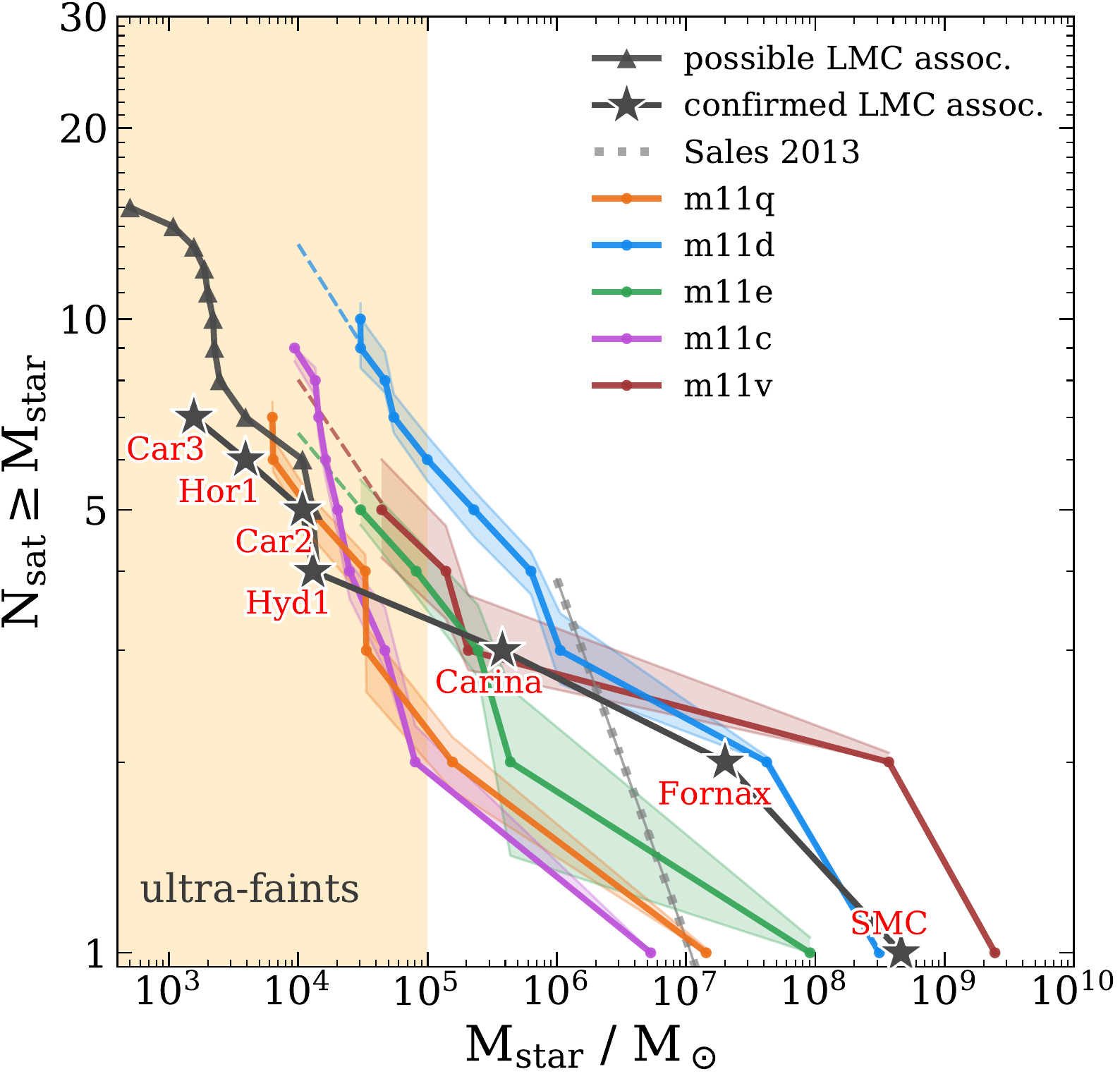}
    \hspace{5mm}
    \includegraphics[width=0.98\columnwidth]{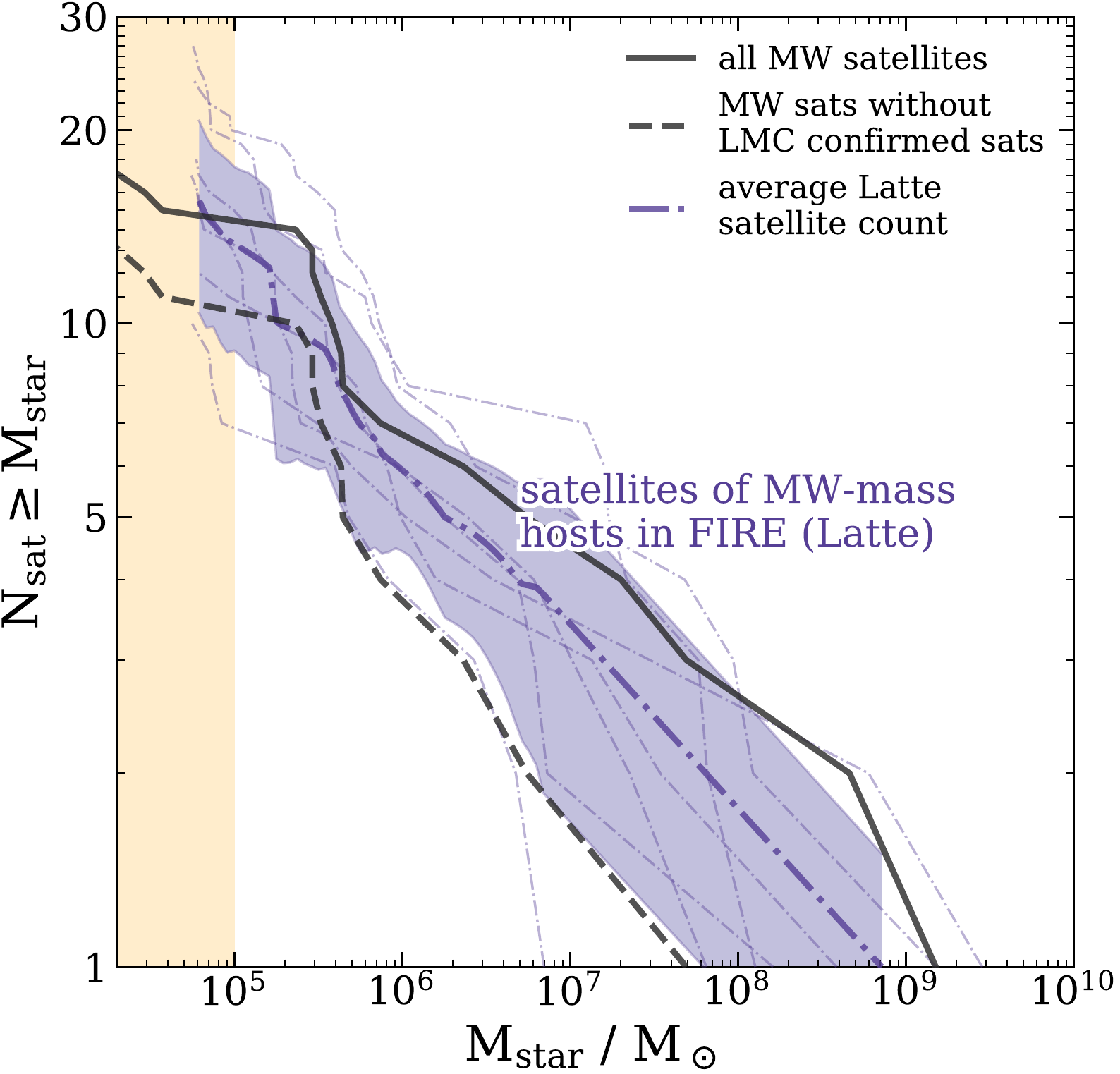}
    \caption{Cumulative $z=0$ count of satellite galaxies above a given stellar mass within one virial radius of the host. (Left) The satellite stellar mass functions of LMC-mass hosts in FIRE (colored) and the real LMC (dark grey). Shaded regions show the 1$\rm \sigma$ variance from over the last $\sim$1.3 Gyr. Confirmed LMC satellites are named in red and shown as star markers, while possible LMC satellites are cumulative with the confirmed population and shown as triangles. `Confirmed LMC assoc.' refers to dwarf galaxies with full proper motion measurements which have angular momenta in agreement with the LMC infall direction, while those labeled `possible' have incomplete proper motion data, but existing data is consistent. The teal dotted line is the expected satellite mass function of an LMC-mass host as predicted by semi-analytic modelling in \citet{sales2013dwfsats}, which uses the model in \citet{guo2011}. All error bars are Poisson noise. (Right) The solid grey line represents all satellite galaxies of the MW, while the dashed grey line represents the same satellites of the MW minus the confirmed satellites of the LMC, which are shown in the left panel and listed in Table \ref{tab:props}. This shows what the in-situ satellite population of the MW was prior to the infall of the LMC and its associated companions. The purple shaded region represents the range of satellite mass functions of these MW-mass hosts in FIRE, with thin lines representing each individual host, and thick line indicating the average number of luminous satellites at each mass. The yellow vertical shaded region on the left represent the ultrafaint mass scale.}
    \label{fig:satMF}
\end{figure*}

With the newest proper motions from Gaia, Carina and Fornax are also compatible with having been accreted as part of the LMC system (`obs' and $t_{1p}$ are consistent with each other within 1 $\rm \sigma$). With stellar masses of 3.8\e{5} \msun and 2.0\e{7} \msuns, respectively, this newly confirmed pair of galaxies fills the classical dwarf scale (10$^5 \lesssim$ \mstrs $\lesssim$ 10$^7$ \msun) in the satellite mass function of the LMC, which was previously populated only by ultrafaint dwarfs (Car2, Car3, Hor1, Hyd1 aside from the from the relatively bright SMC with \mstrs $\sim$ 4.6\e{8} \msuns). 

\citet{pardy2019aurigaLMC} suggest the possibility of LMC co-infall for Carina and Fornax by constraining their projected 2D orbital poles to within 30\textdegree of that of the LMC. We expand on this claim by using a more stringent criteria: matching angular momentum orientation with the one expected for and LMC-like debree at the same position on the sky of Carina and Fornax. Notice that the radial velocities for both dwarfs have been already found consistent with an LMC association in previous work \citep{sales2011MagGal,sales2017truMCsats}. Also worth highlighting, the galactocentric distance of Carina is in good agreement with predictions of association whereas in the case of Fornax, the measured galactocentric distance ($\sim 140$ kpc) places it beyond the $\sim 100$ kpc preferred location of the debree \citep[see for instance Fig.6 in ][]{sales2011MagGal}. The large distance of Fornax is more consistent with the scenario of a more massive infall halo mass for the LMC (whereas previous predictions were based on an LMC-analog with $\sim 10$ times lower mass), which would allow for a more extended distribution of the associated material. This caveat is an important one to bear in mind, and invites further investigation.

Note that Ursa Minor seems to meet the criteria set forth by S17, ($j_x < 0$ and $|j_x| >> |j_y|, |j_z|$). However, its nearly perfect radial orbit ($j_x = -1$, $j_y \approx j_z \approx 0$) as well as its position in a completely distinct region of the sky than predicted for LMC debris (and where all currently known LMC satellites reside; see S17 Figure 1) suggest a non-Magellanic origin for Ursa Minor.

We summarize in Table~\ref{tab:obs} a complete list of MW dwarfs with their current understanding of association to the LMC. The rightmost columns labelled `possible' (if follow up is needed) and `confirmed' (if enough information exists to make the claim) with the relevant references. The previously confirmed satellites of the LMC (by S17 and K18) include Car2, Car3, Hor1, Hyd1 and the SMC. No label means that a given galaxy is unlikely to be associated with the LMC given the current data. Galaxies confirmed by our calculations using Gaia DR2 are labelled `this work'.

\subsection{Simulated LMC Satellite Populations}
With new observational context to the number of dwarf galaxies consistent with co-evolution and co-infall with the LMC, we can examine these results in a cosmological context. We provide this context by analyzing the satellite population of \lcdms cosmological zoom-in simulations of isolated LMC-mass hosts. The left panel in Fig.~\ref{fig:satMF} shows the stellar mass function of LMC satellites in FIRE (colored lines refer to the same simulations as previous figures, with the dashed lines representing an extrapolation to \mstrs $\sim 10^4$ \msuns for the runs with resolution $m_\text{bary}$ = 7070 \msun). In dark gray we show the observed stellar mass function of LMC satellties inferred from the kinematics of MW dwarfs from Gaia DR2 data, using starred symbols for the confirmed associations (SMC, Carina, Fornax, Hyd1, Car2, Hor1, and Car3) and in triangles including all `possible' associations to the LMC, as determined by S17 and K18. 

We find an overall good agreement between the inferred satellite population of the LMC and our simulated analogs. Our simulations predict between 1 and 5 classical satellites of the LMC, in agreement with the observational estimate of 3 for the LMC (SMC, Carina, and Fornax). There is an interesting mass dependence on the ability to predict relatively massive satellites for an LMC-like host. Only the two highest mass FIRE hosts (\texttt{m11d} with \mtwo=2.8\e{11} \msun, and \texttt{m11v} with \mtwo=2.9\e{11} \msun) are able to reproduce the high-mass end of the LMC's satellite mass function, in very close agreement with the halo mass estimates of the LMC ($\sim$3\e{11} \msun) from other methods based on abundance matching \citep{behroozi2013efficiency,moster2013halos} and circular velocity measurements \citep{vanderMarel2014LMC}. In fact, the average halo mass of a LMC-SMC system in the EAGLE simulations is $\sim$3\e{11} \msuns \citep{shao2018lmceagle,cautun2019collision}. The remaining three centrals with halo mass $\sim 1.5$\e{11} \msun, tend to have lower mass companions than the SMC. On the other hand, all runs have at least one satellite within a factor of two the stellar mass of Fornax, supporting its association to the LMC as suggested by the newly released Gaia kinematics.

One should keep in mind that the LMC-SMC association itself is rather unusual. Previous work on LMC-SMC selected pairs have showed them to be rare, though not impossible \citep{boylankolchin2011dynamics, stierwalt2015}. For example, \citet{besla2018dgm} used Illustris and SDSS to predict that the number of companions with \mstr $\sim$ 2\e{8} \msuns per LMC-mass dwarf is roughly 0.02 once projection effects have been taken into account. It is unclear how this figure changes with host mass, but following our results on the trend with virial mass, the likelihood of such a companion should increase if the LMC halo is on the massive end of the 1 $ - $ 3 $\times 10^{11}$ \msuns range\footnote{\texttt{m11v} was specifically selected to be a multimerger, and thus cannot aid a discussion of the cosmological frequency of large satellites}. In general, the number of classical dwarfs in our FIRE centrals is in good agreement with earlier predictions from semi-analytical calculations in \citet[light-blue dotted line]{sales2013dwfsats} taken from the Millennium-II simulations \citep{boylan2009Mil2}. However, the overall slope on the massive end of the FIRE runs is shallower than that in  \citet{sales2013dwfsats}, probably a result of a slightly different stellar mass - halo mass relation for low mass dwarfs in the semi-analytical catalog than in our hydrodynamical runs.

For the ultrafaint regime, our two highest resolution runs, \texttt{m11q} and \texttt{m11c}, predict $5 - 8$ companions with \mstrs $ \geq 10^4$ \msuns, which is in good agreement with the 5 inferred for the LMC from proper motion observations of MW dwarfs \citep{kallivayalil2018MCsats}. In general, if extrapolating the medium resolution runs to \mstrs $ \geq 10^4$ \msuns (which, according to the highest resolution runs, would be a reasonable approach) the predicted number of satellites above this stellar mass may be as high as 11 for \texttt{m11d}. Although such predictions should be taken with caution given the low number statistics and numerical resolution, the assumption that some of the LMC ultrafaint companions still await discovery is a reasonable one. In particular some of the brighter `possible' associations already identified, Hyd2 with \mstrs $ \sim$ 1.4\e{4} \msuns and Dra2 with \mstrs $ \sim$ 2.5\e{3} \msun, might deserve a closer follow up to confirm or rule out their Magellanic origin. 



 Furthermore, our estimates for the number of ultra-faint companions can be regarded as lower limits, because of the implemented model for cosmic-ray heating in the ISM (in addition to the assumed UV background), which in these FIRE simulations induces too much heating in gas at early times ($z \gtrsim 10$) \citep{gk2019sfh}. This effect has been explicitly tested with no observed impact on the population of classical dwarfs, however, the additional heating may in fact \textit{decrease} the number of predicted ultrafaints compared to a method resulting in later reionization \citep[for details, see][Section 3.3 and Appendix B]{gk2019sfh}. As such, our estimates are effectively a conservative lower limit and indicate that several yet-to-be-discovered companions to the LMC may yet exist.

\begin{figure}
    \centering
    \includegraphics[width=0.98\columnwidth]{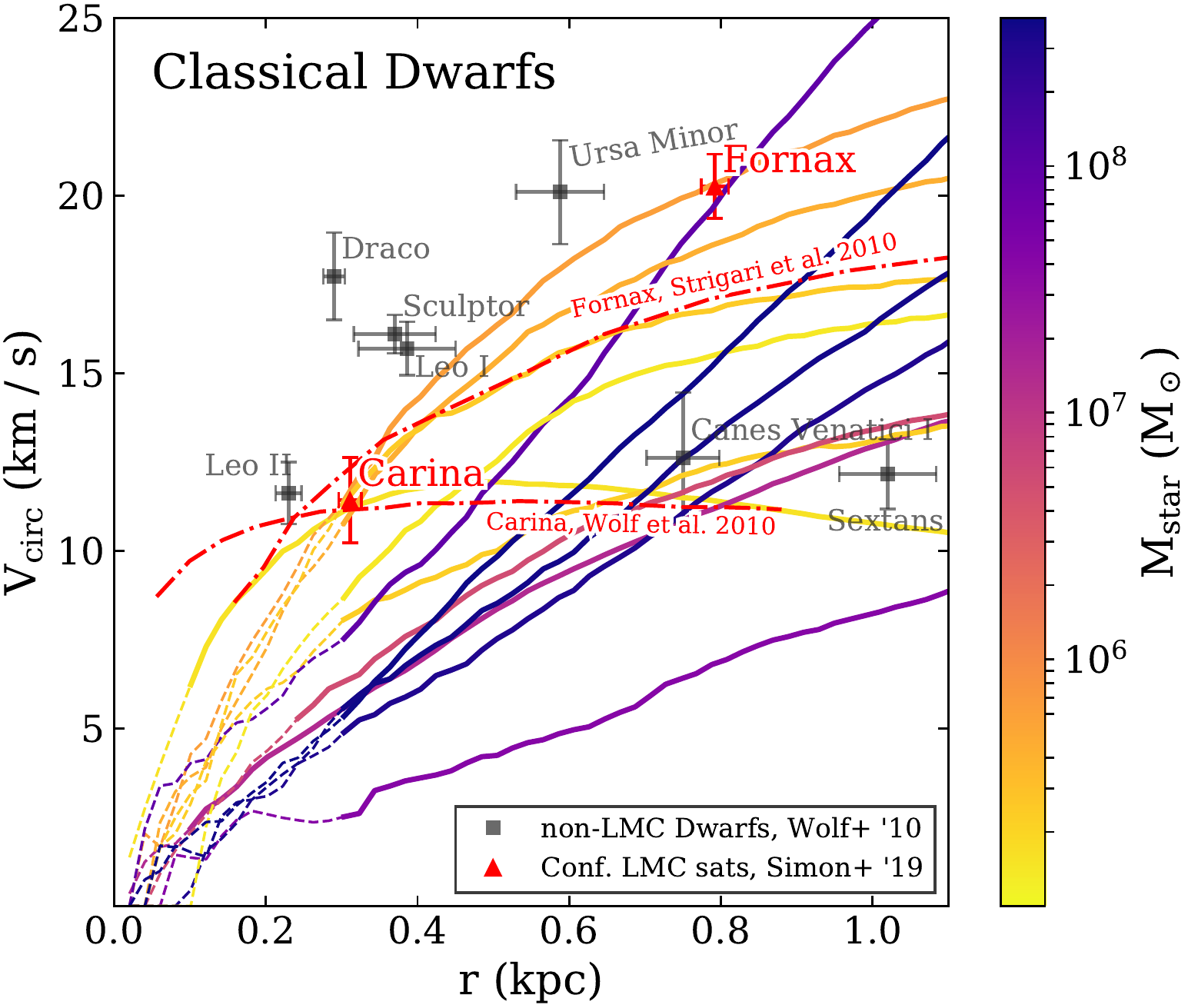}
    \caption{Circular velocity profiles  $\big($\vcirc=$\sqrt{GM(<r)/r}$ $\big)$ of all \texttt{m11} satellites in FIRE. Due to resolution, only satellites with \mstrs above (below) 1\e{5} \msun are shown, colored according to \mstr.  All individual points are observed half-light radii (\textit{x-axis)} versus $V_\text{circ}(r_{1/2}) = \sqrt{3\langle\sigma^2_\text{los}\rangle}$ as described in \citet{boylankolchin2012MWsats}. Grey points (taken from \citealt{wolf2010masses}) are various MW dwarfs that do not belong to the LMC-system, while those dwarf galaxies presumably associated with the LMC are painted in red. The two red lines are circular velocities for Carina and Fornax as determined by \citet{strigari2010kinematics} and \citet{wolf2010masses}, noted accordingly. The dark matter content of our predicted classical satellites is in good agreement with the newly deemed members, Carina and Fornax.} 
    \label{fig:subhalo_vcirc}
\end{figure}

Lastly, on the right panel of Fig.~\ref{fig:satMF} we show the observed stellar mass function of MW satellites (black solid) along with satellites of M31 (black dot dashed). If the LMC brought along several of the dwarfs as estimated from the previous calculations (gray starred symbols), the MW satellite mass function must have looked rather different about 1 Gyr ago right before the infall of the LMC.\footnote{However, exact counts of luminous satellites are subject to change over this timescale as they are disrupted into streams, e.g. Sagittarius.} This is shown as the gray dashed line, computed as the total MW satellites but subtracting the confirmed LMC associations. As usual, we define satellites as those within \rtwos of the host. In such case, the MW halo may have hosted a significantly lower number of dwarfs than now, although still in reasonable agreement with the predicted satellite population of Latte galaxies (\mtwos $ \sim 10^{12}$ \msun, shaded in purple). The similarity between the high-mass end of the satellite mass function for the MW and the LMC argues once again for a rather massive pre-infall LMC halo, likely $\sim 3 \times 10^{11}$ \msuns and above, predicting several undiscovered ultra-faint dwarfs that were previously associated to the LMC.

\subsection{The Predicted Dark Matter Content of LMC Satellites}
\label{ssec:dmcontent}

Besides the number of dwarf galaxies expected around Magellanic-like systems, a further (arguably stronger) test for \lcdms galaxy formation models is to reproduce the internal kinematics of the stars that are measured from observations. In the case of ultrafaint dwarfs, reported velocities from observations cover mostly the radius range $r<200$ pc; with at least half of the systems having measured velocities within 50 pc \citep{simon2019UFDs}. Unfortunately, integrated quantities such as circular velocities are not yet converged in our simulations at such extreme small radii (typical gravitational softening $\epsilon_\text{\tiny DM}~\sim 20 - 40$ pc for all runs). We therefore analyze kinematic profiles of simulated classical dwarfs, but present a study of \vmaxs that can guide the conclusions in the ultrafaint regime.

Fig.~\ref{fig:subhalo_vcirc} shows the circular velocity profiles of all simulated classical satellites (\mstrs > $10^5$ \msun) of our FIRE LMC-analogs. Lines are dashed below the dark matter convergence radius for each resolution as listed in \citet{hopkins2018fire2}. Individual curves are color coded according to the stellar mass content of each satellite, as indicated by the color bar. The simulation data hints to a correlation where the larger \mstrs dwarfs will populate larger circular velocity subhalos, at least in the regime of classical dwarfs.  

\begin{figure}
    \centering
    \includegraphics[width=\columnwidth]{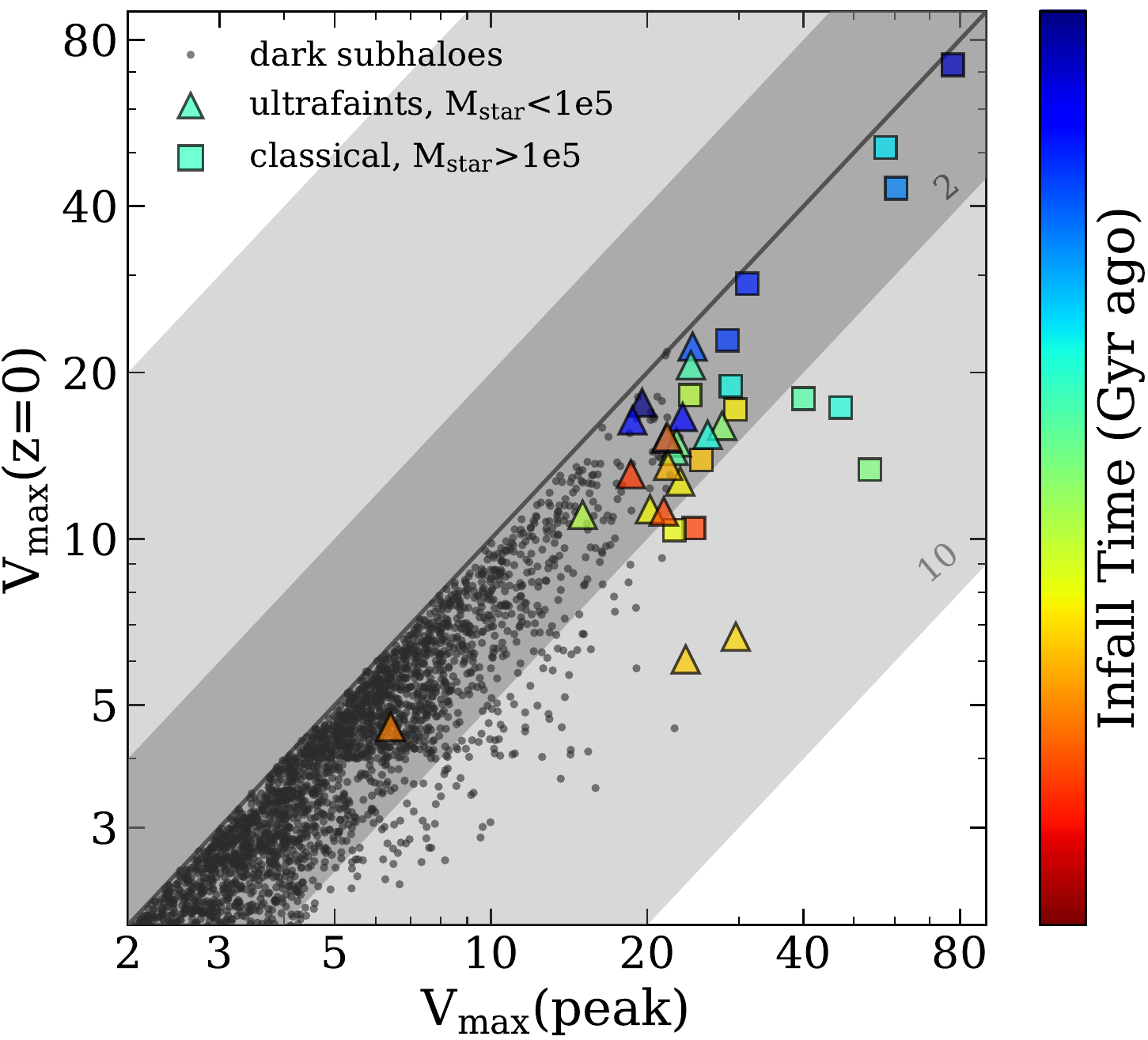}
    \caption{The peak \vmaxs ever obtained by the subhalo versus its present-day \vmax. Squares represent subhalos which host galaxies of \mstr$\geq$10$^\text{5}$ \msun, triangles host those with \mstr$\leq$10$^\text{5}$ \msun, and grey dots are dark subhalos. The color bar shows the time in Gyr since a given satellite's infall to the host LMC-mass central. As expected, satellites with recent infall times ($T_\text{inf} \lesssim 4$ Gyr) show minimal deviation from $V_\text{peak}$. However, tidal disruption within LMC-mass systems for earlier infalling satellites may cause a factor of $\sim 2-7$ reduction in the \vmaxs of ultrafaints at $z=0$.}
    \label{fig:vpeak}
\end{figure}

Observed dwarf spheroidals are also plotted as single points assuming \vcirc($r_{1/2}$) = $\sqrt{3\langle\sigma_\text{los}\rangle}$ with $\sigma_{\rm los}$ the observed line-of-sight stellar velocity on the $y$-axis and the half-light radii $r_{1/2}$ on the $x$-axis (following \citealt{boylankolchin2012MWsats}). Sources for the observed $\sigma_{v,\ast}$ are listed in the legend of Fig.~\ref{fig:subhalo_vcirc}. Observed galaxies presumably associated with the LMC are shown in red while other MW dwarfs are plotted in gray. For completeness, we also include the full circular velocity profiles as a function of radius for Carina and Fornax from \citet{wolf2010masses} and \citet{strigari2010kinematics}, respectively. 

 The predicted mass content for the brightest satellites of LMC-mass hosts, such as Carina and Fornax, are consistent with the observational values, supporting the possible association to the LMC inferred from their proper motions. It is important to highlight that the simulated curves in Fig.~\ref{fig:subhalo_vcirc} are, if anything, lower limits to the true density of simulated dwarfs, as a smaller softening and increased resolution would result in higher inner velocities,\footnote{We have tested this explicitly, see Fig.~\ref{fig:vcirc_restest}. Also see \citet{Springel08Aq} for an evaluation of softening effects on \vcirc.} which would still accommodate the observed values. The good agreement with observations of Carina and Fornax is therefore encouraging and highly suggestive of a possible membership to the LMC group.

For the ultrafaint regime, we turn our analysis to maximum circular velocities, as we expect them to be well converged. Following Eq. 10 in \citet{Springel08Aq}, the correction due to numerical effects is at most $\sim$0.1 km/s, or $\lesssim$1\%.  
Interestingly, a closer inspection of our simulations reveals that tidal stripping is likely to play a major role on the present-day dark matter content of LMC companions, particularly in the ultrafaint regime. This is clearly seen in Fig.~\ref{fig:vpeak}, which shows a comparison between peak maximum circular velocity achieved throughout a subhalo's history ($V_{\rm peak}$) as a function of the circular velocity measured at present day (\vmax$(z=0)$) for all subhalos in our sample of LMC-mass hosts in FIRE. Dark subhalos are indicated with gray dots, while large squares and triangles highlight the location of classical and ultrafaint simulated dwarfs, respectively. The symbols are color coded by lookback time since accretion (lower values are more recent infalls), and confirm that some of the surviving ultrafaint satellites in an LMC-system may have been accreted as early as 12 Gyr ago. The two regions of dark and light shaded gray indicate factors of 2 and 10 decrease in circular velocity.

As indicated by Fig.~\ref{fig:vpeak}, the ultrafaint LMC companions are narrowly distributed at \vpeaks $\sim$ 20 km/s at infall but show a large spread in \vcircs today, 6 $ - $ 20 km/s. 
We find the present-day median \vmaxs for ultrafaint LMCs to be 14.4 km/s. As expected, the latest ones to infall remain close to the 1-to-1 line, as tidal disruption has not had sufficient time to affect their properties. We find that he amount of tidal stripping experienced is not dependent on stellar mass. However, consider $(i)$ the narrow range of \vpeaks predicted for ultrafaints, and $(ii)$ their low stellar masses, at which feedback is not expected to affect the inner DM distribution. Both effects combined establish an interesting correlation in ultrafaints that might be used to assess their likelihood of association to the LMC. We emphasize that low central densities are a necessary rather than sufficient condition to determine an association to the LMC since an early infall onto the MW for ultrafaints will also induce tidal stripping and associated lower inner densities. Determining which host a satellite was first associated with is a task left to orbital phase-space analysis, as in sub-section \ref{ssec:gaia}.

Although tidal stripping proceeds mostly outside-in, subhalos that are affected by tides also register a drop in their inner dark matter content \citep[e.g., ][]{Hayashi2003, brooks2014dpshkin}. It is therefore expected that ultrafaint satellites of the LMC might show lower inner velocities than objects of similar mass that were accreted more recently into the MW. This prospect is interesting since Gaia DR2 data suggests that several of the ultrafaint galaxies are likely on their first infall onto the MW \citep{simon2018GAIA}. Identifying ultrafaint dwarfs with orbital properties consistent with that of the LMC and that simultaneously show the lowest inner densities may help to constrain the most likely LMC group members\footnote{Note that the MW environment may also be conducive to tidal stripping such that low inner densities may be a telling but not sufficient condition for association to the LMC.}. From that perspective, the likely associations for Hyd1 with \vcircs = 4.7 km/s and Car2 with \vcircs = 5.9 km/s, both within $r\leq 100$ pc, or the possibly associated Hyd2 with \vcircs = 6.2 km/s, seem favored compared to the more dense Car3, Hor1, or Dra2 with \vcircs $\geq 8$ km/s at $r\sim$ 50 pc \citep[data from ][]{simon2019UFDs}. However, higher resolution simulations as well as more accurate observations are needed in order to make more definitive claims.

\section{Summary and Conclusions}
\label{sec:Conclusion}

We used five \lcdms cosmological zoom in simulations of LMC-mass hosts  (\mtwos$ = 1 - 3 \times 10^{11}$ \msun) in FIRE to examine both the dark and luminous substructure of such galaxies, against which we compare to the Latte suite of seven simulations of MW-mass hosts. We summarize our primary findings here, and discuss our conclusions in the following paragraphs.
\begin{enumerate}
    \item By comparing dark matter only to hydrodynamic baryonic simulations, we show that suppression of dark matter substructure is less strong for LMC-mass hosts than for MW-mass hosts, since this suppression is caused by tidal interactions with the central baryonic galaxy \citep{notsolumpy}. We therefore expect that LMC-mass galaxies are promising laboratories for subhalo detection.
    \item We calculate orbital angular momenta for 10 observed MW-dwarfs using Gaia DR2 data \citep{helmi2018gaia} and show that Fornax and Carina are highly consistent with predictions for tidal debris of a simulated LMC-MW infall scenario from \citet{sales2017truMCsats}. This brings the inferred satellite mass function of the LMC up to seven members, including the SMC and the four ultrafaints identified in \citet{kallivayalil2018MCsats}.
    \item We compare this to the simulated satellite mass function of our five LMC-mass hosts in FIRE and find excellent agreement on the bright end with our higher mass halos. Our simulations suggest that more ultrafaints are expected for a halo comparable to that of the LMC. In addition, the population of MW satellites pre-LMC infall remains consistent with simulated MW-mass hosts in FIRE.
    \item We find that the tidal disruption of simulated LMC satellites, indicated by a reduction in their maximum circular velocities after infall, is an important effect, even at such host mass scales. We therefore expect that dwarfs associated with the LMC should have lowered densities, though this is not sufficient criteria for association by itself as tidal effects from the MW are also expected to affect nearby dwarfs.  
\end{enumerate}

We have shown that LMC-mass centrals suppress much less dark matter substructure compared to MW-mass centrals, with a particularly striking difference in suppression patterns for the inner 0.2 \rtwos of the halo. There, we observe a factor of $\sim$3.5 reduction (from dark matter only to baryonic runs) in the number of subhalos at \vmaxs = 10 km/s in MW mass hosts versus a factor of $\sim$1.4 for LMC mass hosts. This suppression of substructure is mainly due to tidal interactions with the central baryonic disk. We identify two features of the LMC-mass hosts in FIRE that explains this: (i) a much lower stellar mass to halo mass ratio than that of MW-mass objects and (ii) a significantly shallower inner density profile (larger core). The first implies that the gravitational potential of the galaxy is less significant compared to that of the halo. The second implies that the shape of the potential is much smoother than that seen in DMO versions of similar halos and that seen in halos hosting larger central galaxies. This combination provides a friendlier environment in the baryonic runs of cored \mtwos $ \sim 10^{11}$ \msuns halos, leading to less suppression of subhalos due to tidal disruption on these scales compared to the harsher effects reported for $10^{12}$ \msuns hosts \citep{donghia2010,notsolumpy,samuel2019radial}.

We compare the simulated satellite population of the LMC-mass hosts to the set of observed MW dwarfs that is consistent with an LMC co-infall scenario. We present revised calculations following \citet{sales2011MagGal} and using updated Gaia DR2 proper motion for classical dwarfs  (\mstrs $\geq 10^5$ \msun) from \citet{helmi2018gaia}. We find that Carina and Fornax are now compatible with a common infall along with the LMC system, and are added to the previously suggested associations to the SMC and the group of ultrafaint candidates Car2, Car3, Hor1, and Hyd1 introduced in \citet{kallivayalil2018MCsats}. 
We find generally good agreement with the satellite population inferred for the real LMC. One of our halos, \texttt{m11d} with \mtwos $ \sim 3 \times 10^{11}$ \msun, is able to accurately predict the mass distribution of the three largest LMC satellites: the SMC, Fornax, and Carina, providing theoretical support to the claims of association. Furthermore, the predicted circular velocities for LMC satellites in this mass range is in good agreement with measurements of Carina and Fornax. 

For fainter satellites, the LMC-mass systems in FIRE host comparable to slightly more ultrafaint companions than observed with the real LMC. On average, $7\pm 2$ satellites are predicted above \mstrs $=10^4$\msun, in reasonable agreement with the 5 above that mass limit presumed associated to the LMC (Car2, Hyd1, Carina, Fornax, and the SMC). We find that ultrafaint companions of the LMC are expected to have experienced significant tidal disruption within the LMC potential, as measured by the decrease on the subhalo maximum circular velocities since infall. As such, lower dark matter inner densities together with their orbital parameters \citep{kallivayalil2018MCsats}, may help identify those ultra faint dwarfs that infalled onto the MW as part of the LMC group. 

In summary, if the SMC, Carina, and Fornax are former satellites of the LMC, this may favor a relatively massive dark matter halo mass for the LMC prior to infall onto the MW, \mtwos $ \sim 3\times 10^{11}$ \msun. This would push the expected ultrafaint dwarf numbers to the upper end of our predicted range, suggesting that some LMC associations still await discovery. According to our simulations, the missing dwarfs will lie roughly $30-80$ kpc from the LMC at infall and have \mstrs $ \sim 10^4$ \msuns and $V_{\rm max} \approx 15$ km/s. This implies a relatively low central dark matter density that can be used as an additional membership criteria to discern from other, more recent individual infalls onto the MW. Based on their partially known orbital parameters and their low velocity estimates, the most promising candidate is Hyd2. Follow up observations are needed to confirm or dismiss the association.  

In addition to LMC, this work presents the first observational and testable predictions using hydrodynamical cosmological simulations for the satellite mass function of a Magellanic-mass system down to the ultrafaint regime (semi-analytic modelling has been used for similar predictions, e.g. \citealt{dooley2017LMCsats,bose2018LF}). Similar to how the study of satellite population of MW-mass hosts pushed forward our understanding of galaxy formation and cosmology in the past, the large predicted number of isolated dwarfs combined with upcoming deep surveys and wide field-of-view instruments such as WFIRST, as well as searches for satellites of LMC-mass hosts beyond the MW \citep[e.g. MADCASH;][]{carlin2016madcash}, may turn the study of dwarf-dwarf systems into a valuable and essential test of the $\Lambda$CDM model.

\section*{Acknowledgments}
The authors would like to thank Phil Hopkins for facilitating access to the FIRE runs and stimulating discussions of the simulations and science herein. We also thank Coral Wheeler, Shea Garrison-Kimmel, and the rest of the FIRE local universe collaboration for their thought-provoking insight and advice. We are grateful to the anonymous referee for a constructive report that helped improve this manuscript, as well as to Marius Cautun and Marcel Pawlowski for their helpful suggestions. 
LVS acknowledges support from NASA through the HST Programs AR-14582, AR-14583 and 
from the Hellman Foundation.
AW was supported by NASA, through ATP grant 80NSSC18K1097 and HST grants GO-14734 and AR-15057 from STScI.
MBK acknowledges support from NSF grant AST-1517226 and CAREER grant AST-1752913 and from NASA grants NNX17AG29G and HST-AR-14282, HST-AR-14554, HST-AR-15006, and HST-GO-14191 from the Space Telescope Science Institute, which is operated by AURA, Inc., under NASA contract NAS5-26555.
TKC was supported by NSF grant AST-1715101 and the Cottrell Scholar Award from the Research Corporation for Science Advancement.
We ran simulations using XSEDE supported by NSF grant ACI-1548562, Blue Waters via allocation PRAC NSF.1713353 supported by the NSF, and NASA HEC Program through the NAS Division at Ames Research Center.

\bibliographystyle{mnras}
\bibliography{lmc_paper}

\begin{thebibliography}{}
\makeatletter
\relax
\def\mn@urlcharsother{\let\do\@makeother \do\$\do\&\do\#\do\^\do\_\do\%\do\~}
\def\mn@doi{\begingroup\mn@urlcharsother \@ifnextchar [ {\mn@doi@}
  {\mn@doi@[]}}
\def\mn@doi@[#1]#2{\def\@tempa{#1}\ifx\@tempa\@empty \href
  {http://dx.doi.org/#2} {doi:#2}\else \href {http://dx.doi.org/#2} {#1}\fi
  \endgroup}
\def\mn@eprint#1#2{\mn@eprint@#1:#2::\@nil}
\def\mn@eprint@arXiv#1{\href {http://arxiv.org/abs/#1} {{\tt arXiv:#1}}}
\def\mn@eprint@dblp#1{\href {http://dblp.uni-trier.de/rec/bibtex/#1.xml}
  {dblp:#1}}
\def\mn@eprint@#1:#2:#3:#4\@nil{\def\@tempa {#1}\def\@tempb {#2}\def\@tempc
  {#3}\ifx \@tempc \@empty \let \@tempc \@tempb \let \@tempb \@tempa \fi \ifx
  \@tempb \@empty \def\@tempb {arXiv}\fi \@ifundefined
  {mn@eprint@\@tempb}{\@tempb:\@tempc}{\expandafter \expandafter \csname
  mn@eprint@\@tempb\endcsname \expandafter{\@tempc}}}

\bibitem[\protect\citeauthoryear{{Abadi}, {Navarro}, {Steinmetz}  \&
  {Eke}}{{Abadi} et~al.}{2003}]{abadi2003sims}
{Abadi} M.~G.,  {Navarro} J.~F.,  {Steinmetz} M.,   {Eke} V.~R.,  2003, \mn@doi
  [\apj] {10.1086/375512}, \href
  {https://ui.adsabs.harvard.edu/abs/2003ApJ...591..499A} {591, 499}

\bibitem[\protect\citeauthoryear{{Bechtol} et~al.,}{{Bechtol}
  et~al.}{2015}]{bechtol2015DES}
{Bechtol} K.,  et~al., 2015, \mn@doi [\apj] {10.1088/0004-637X/807/1/50}, \href
  {https://ui.adsabs.harvard.edu/abs/2015ApJ...807...50B} {807, 50}

\bibitem[\protect\citeauthoryear{{Behroozi}, {Conroy}  \&
  {Wechsler}}{{Behroozi} et~al.}{2010}]{behroozi10smhm}
{Behroozi} P.~S.,  {Conroy} C.,   {Wechsler} R.~H.,  2010, \mn@doi [\apj]
  {10.1088/0004-637X/717/1/379}, \href
  {http://adsabs.harvard.edu/abs/2010ApJ...717..379B} {717, 379}

\bibitem[\protect\citeauthoryear{{Behroozi}, {Wechsler}  \& {Wu}}{{Behroozi}
  et~al.}{2013a}]{behroozi2012rockstar}
{Behroozi} P.~S.,  {Wechsler} R.~H.,   {Wu} H.-Y.,  2013a, \mn@doi [\apj]
  {10.1088/0004-637X/762/2/109}, \href
  {https://ui.adsabs.harvard.edu/abs/2013ApJ...762..109B} {762, 109}

\bibitem[\protect\citeauthoryear{{Behroozi}, {Wechsler}  \&
  {Conroy}}{{Behroozi} et~al.}{2013b}]{behroozi2013efficiency}
{Behroozi} P.~S.,  {Wechsler} R.~H.,   {Conroy} C.,  2013b, \mn@doi [\apj]
  {10.1088/0004-637X/770/1/57}, \href
  {https://ui.adsabs.harvard.edu/abs/2013ApJ...770...57B} {770, 57}

\bibitem[\protect\citeauthoryear{{Besla} et~al.,}{{Besla}
  et~al.}{2018}]{besla2018dgm}
{Besla} G.,  et~al., 2018, \mn@doi [\mnras] {10.1093/mnras/sty2041}, \href
  {https://ui.adsabs.harvard.edu/abs/2018MNRAS.480.3376B} {480, 3376}

\bibitem[\protect\citeauthoryear{{Bose}, {Deason}  \& {Frenk}}{{Bose}
  et~al.}{2018}]{bose2018LF}
{Bose} S.,  {Deason} A.~J.,   {Frenk} C.~S.,  2018, \mn@doi [\apj]
  {10.3847/1538-4357/aacbc4}, \href
  {https://ui.adsabs.harvard.edu/abs/2018ApJ...863..123B} {863, 123}

\bibitem[\protect\citeauthoryear{{Boylan-Kolchin}, {Springel}, {White},
  {Jenkins}  \& {Lemson}}{{Boylan-Kolchin} et~al.}{2009}]{boylan2009Mil2}
{Boylan-Kolchin} M.,  {Springel} V.,  {White} S. D.~M.,  {Jenkins} A.,
  {Lemson} G.,  2009, \mn@doi [\mnras] {10.1111/j.1365-2966.2009.15191.x},
  \href {https://ui.adsabs.harvard.edu/abs/2009MNRAS.398.1150B} {398, 1150}

\bibitem[\protect\citeauthoryear{{Boylan-Kolchin}, {Besla}  \&
  {Hernquist}}{{Boylan-Kolchin} et~al.}{2011a}]{boylankolchin2011dynamics}
{Boylan-Kolchin} M.,  {Besla} G.,   {Hernquist} L.,  2011a, \mn@doi [\mnras]
  {10.1111/j.1365-2966.2011.18495.x}, \href
  {http://adsabs.harvard.edu/abs/2011MNRAS.414.1560B} {414, 1560}

\bibitem[\protect\citeauthoryear{{Boylan-Kolchin}, {Bullock}  \&
  {Kaplinghat}}{{Boylan-Kolchin} et~al.}{2011b}]{boylan2011tbtf}
{Boylan-Kolchin} M.,  {Bullock} J.~S.,   {Kaplinghat} M.,  2011b, \mn@doi
  [\mnras] {10.1111/j.1745-3933.2011.01074.x}, \href
  {https://ui.adsabs.harvard.edu/abs/2011MNRAS.415L..40B} {415, L40}

\bibitem[\protect\citeauthoryear{{Boylan-Kolchin}, {Bullock}  \&
  {Kaplinghat}}{{Boylan-Kolchin} et~al.}{2012}]{boylankolchin2012MWsats}
{Boylan-Kolchin} M.,  {Bullock} J.~S.,   {Kaplinghat} M.,  2012, \mn@doi
  [\mnras] {10.1111/j.1365-2966.2012.20695.x}, \href
  {http://adsabs.harvard.edu/abs/2012MNRAS.422.1203B} {422, 1203}

\bibitem[\protect\citeauthoryear{{Brooks} \& {Zolotov}}{{Brooks} \&
  {Zolotov}}{2014}]{brooks2014dpshkin}
{Brooks} A.~M.,  {Zolotov} A.,  2014, \mn@doi [\apj]
  {10.1088/0004-637X/786/2/87}, \href
  {https://ui.adsabs.harvard.edu/abs/2014ApJ...786...87B} {786, 87}

\bibitem[\protect\citeauthoryear{{Brooks}, {Kuhlen}, {Zolotov}  \&
  {Hooper}}{{Brooks} et~al.}{2013}]{brooks2013}
{Brooks} A.~M.,  {Kuhlen} M.,  {Zolotov} A.,   {Hooper} D.,  2013, \mn@doi
  [\apj] {10.1088/0004-637X/765/1/22}, \href
  {http://adsabs.harvard.edu/abs/2013ApJ...765...22B} {765, 22}

\bibitem[\protect\citeauthoryear{{Bryan} \& {Norman}}{{Bryan} \&
  {Norman}}{1998}]{bryannorman1998}
{Bryan} G.~L.,  {Norman} M.~L.,  1998, \mn@doi [\apj] {10.1086/305262}, \href
  {https://ui.adsabs.harvard.edu/abs/1998ApJ...495...80B} {495, 80}

\bibitem[\protect\citeauthoryear{{Bullock} \& {Boylan-Kolchin}}{{Bullock} \&
  {Boylan-Kolchin}}{2017}]{bullock2017lcdm}
{Bullock} J.~S.,  {Boylan-Kolchin} M.,  2017, \mn@doi [\araa]
  {10.1146/annurev-astro-091916-055313}, \href
  {https://ui.adsabs.harvard.edu/abs/2017ARA&A..55..343B} {55, 343}

\bibitem[\protect\citeauthoryear{{Bullock}, {Kravtsov}  \&
  {Weinberg}}{{Bullock} et~al.}{2000}]{bullock2000reionabund}
{Bullock} J.~S.,  {Kravtsov} A.~V.,   {Weinberg} D.~H.,  2000, \mn@doi [\apj]
  {10.1086/309279}, \href {http://adsabs.harvard.edu/abs/2000ApJ...539..517B}
  {539, 517}

\bibitem[\protect\citeauthoryear{{Carlberg}, {Grillmair}  \&
  {Hetherington}}{{Carlberg} et~al.}{2012}]{carlberg2012pal5}
{Carlberg} R.~G.,  {Grillmair} C.~J.,   {Hetherington} N.,  2012, \mn@doi
  [\apj] {10.1088/0004-637X/760/1/75}, \href
  {http://adsabs.harvard.edu/abs/2012ApJ...760...75C} {760, 75}

\bibitem[\protect\citeauthoryear{{Carlin} et~al.,}{{Carlin}
  et~al.}{2016}]{carlin2016madcash}
{Carlin} J.~L.,  et~al., 2016, \mn@doi [\apj] {10.3847/2041-8205/828/1/L5},
  \href {https://ui.adsabs.harvard.edu/abs/2016ApJ...828L...5C} {828, L5}

\bibitem[\protect\citeauthoryear{{Cautun}, {Deason}, {Frenk}  \&
  {McAlpine}}{{Cautun} et~al.}{2019}]{cautun2019collision}
{Cautun} M.,  {Deason} A.~J.,  {Frenk} C.~S.,   {McAlpine} S.,  2019, \mn@doi
  [\mnras] {10.1093/mnras/sty3084}, \href
  {https://ui.adsabs.harvard.edu/abs/2019MNRAS.483.2185C} {483, 2185}

\bibitem[\protect\citeauthoryear{{Chan}, {Kere{\v{s}}}, {O{\~n}orbe},
  {Hopkins}, {Muratov}, {Faucher-Gigu{\`e}re}  \& {Quataert}}{{Chan}
  et~al.}{2015}]{chan2015FIREcores}
{Chan} T.~K.,  {Kere{\v{s}}} D.,  {O{\~n}orbe} J.,  {Hopkins} P.~F.,  {Muratov}
  A.~L.,  {Faucher-Gigu{\`e}re} C.~A.,   {Quataert} E.,  2015, \mn@doi [\mnras]
  {10.1093/mnras/stv2165}, \href
  {https://ui.adsabs.harvard.edu/abs/2015MNRAS.454.2981C} {454, 2981}

\bibitem[\protect\citeauthoryear{{Chan}, {Kere{\v{s}}}, {Wetzel}, {Hopkins},
  {Faucher-Gigu{\`e}re}, {El-Badry}, {Garrison-Kimmel}  \&
  {Boylan-Kolchin}}{{Chan} et~al.}{2018}]{chan2018udgs}
{Chan} T.~K.,  {Kere{\v{s}}} D.,  {Wetzel} A.,  {Hopkins} P.~F.,
  {Faucher-Gigu{\`e}re} C.~A.,  {El-Badry} K.,  {Garrison-Kimmel} S.,
  {Boylan-Kolchin} M.,  2018, \mn@doi [\mnras] {10.1093/mnras/sty1153}, \href
  {https://ui.adsabs.harvard.edu/abs/2018MNRAS.478..906C} {478, 906}

\bibitem[\protect\citeauthoryear{{Chiba}}{{Chiba}}{2002}]{chiba2002dmsubstructure}
{Chiba} M.,  2002, \mn@doi [\apj] {10.1086/324493}, \href
  {http://adsabs.harvard.edu/abs/2002ApJ...565...17C} {565, 17}

\bibitem[\protect\citeauthoryear{{D'Onghia} \& {Lake}}{{D'Onghia} \&
  {Lake}}{2008}]{donghialake08}
{D'Onghia} E.,  {Lake} G.,  2008, \mn@doi [\apj] {10.1086/592995}, \href
  {https://ui.adsabs.harvard.edu/abs/2008ApJ...686L..61D} {686, L61}

\bibitem[\protect\citeauthoryear{{D'Onghia}, {Springel}, {Hernquist}  \&
  {Keres}}{{D'Onghia} et~al.}{2010}]{donghia2010}
{D'Onghia} E.,  {Springel} V.,  {Hernquist} L.,   {Keres} D.,  2010, \mn@doi
  [\apj] {10.1088/0004-637X/709/2/1138}, \href
  {https://ui.adsabs.harvard.edu/abs/2010ApJ...709.1138D} {709, 1138}

\bibitem[\protect\citeauthoryear{{Deason}, {Wetzel}, {Garrison-Kimmel}  \&
  {Belokurov}}{{Deason} et~al.}{2015}]{deason2015LMCsats}
{Deason} A.~J.,  {Wetzel} A.~R.,  {Garrison-Kimmel} S.,   {Belokurov} V.,
  2015, \mn@doi [\mnras] {10.1093/mnras/stv1939}, \href
  {https://ui.adsabs.harvard.edu/abs/2015MNRAS.453.3568D} {453, 3568}

\bibitem[\protect\citeauthoryear{{Di Cintio}, {Brook}, {Dutton}, {Macci{\`o}},
  {Stinson}  \& {Knebe}}{{Di Cintio} et~al.}{2014}]{dicintio2014profile}
{Di Cintio} A.,  {Brook} C.~B.,  {Dutton} A.~A.,  {Macci{\`o}} A.~V.,
  {Stinson} G.~S.,   {Knebe} A.,  2014, \mn@doi [\mnras]
  {10.1093/mnras/stu729}, \href
  {https://ui.adsabs.harvard.edu/abs/2014MNRAS.441.2986D} {441, 2986}

\bibitem[\protect\citeauthoryear{{Dooley}, {Peter}, {Carlin}, {Frebel},
  {Bechtol}  \& {Willman}}{{Dooley} et~al.}{2017}]{dooley2017LMCsats}
{Dooley} G.~A.,  {Peter} A. H.~G.,  {Carlin} J.~L.,  {Frebel} A.,  {Bechtol}
  K.,   {Willman} B.,  2017, \mn@doi [\mnras] {10.1093/mnras/stx2001}, \href
  {https://ui.adsabs.harvard.edu/abs/2017MNRAS.472.1060D} {472, 1060}

\bibitem[\protect\citeauthoryear{{Drlica-Wagner} et~al.,}{{Drlica-Wagner}
  et~al.}{2015}]{drlica2015DES}
{Drlica-Wagner} A.,  et~al., 2015, \mn@doi [\apj]
  {10.1088/0004-637X/813/2/109}, \href
  {https://ui.adsabs.harvard.edu/abs/2015ApJ...813..109D} {813, 109}

\bibitem[\protect\citeauthoryear{{El-Badry}, {Wetzel}, {Geha}, {Hopkins},
  {Kere{\v{s}}}, {Chan}  \& {Faucher-Gigu{\`e}re}}{{El-Badry}
  et~al.}{2016}]{keb2016feedback}
{El-Badry} K.,  {Wetzel} A.,  {Geha} M.,  {Hopkins} P.~F.,  {Kere{\v{s}}} D.,
  {Chan} T.~K.,   {Faucher-Gigu{\`e}re} C.-A.,  2016, \mn@doi [\apj]
  {10.3847/0004-637X/820/2/131}, \href
  {https://ui.adsabs.harvard.edu/abs/2016ApJ...820..131E} {820, 131}

\bibitem[\protect\citeauthoryear{{El-Badry}, {Wetzel}, {Geha}, {Quataert},
  {Hopkins}, {Kere{\v{s}}}, {Chan}  \& {Faucher-Gigu{\`e}re}}{{El-Badry}
  et~al.}{2017}]{elbadry2017feedback}
{El-Badry} K.,  {Wetzel} A.~R.,  {Geha} M.,  {Quataert} E.,  {Hopkins} P.~F.,
  {Kere{\v{s}}} D.,  {Chan} T.~K.,   {Faucher-Gigu{\`e}re} C.-A.,  2017,
  \mn@doi [\apj] {10.3847/1538-4357/835/2/193}, \href
  {https://ui.adsabs.harvard.edu/abs/2017ApJ...835..193E} {835, 193}

\bibitem[\protect\citeauthoryear{{El-Badry} et~al.,}{{El-Badry}
  et~al.}{2018a}]{keb2018angmom}
{El-Badry} K.,  et~al., 2018a, \mn@doi [\mnras] {10.1093/mnras/stx2482}, \href
  {https://ui.adsabs.harvard.edu/abs/2018MNRAS.473.1930E} {473, 1930}

\bibitem[\protect\citeauthoryear{{El-Badry} et~al.,}{{El-Badry}
  et~al.}{2018b}]{keb2018obskin}
{El-Badry} K.,  et~al., 2018b, \mn@doi [\mnras] {10.1093/mnras/sty730}, \href
  {https://ui.adsabs.harvard.edu/abs/2018MNRAS.477.1536E} {477, 1536}

\bibitem[\protect\citeauthoryear{{Escala} et~al.,}{{Escala}
  et~al.}{2018}]{escala2018metaldiffusion}
{Escala} I.,  et~al., 2018, \mn@doi [\mnras] {10.1093/mnras/stx2858}, \href
  {https://ui.adsabs.harvard.edu/abs/2018MNRAS.474.2194E} {474, 2194}

\bibitem[\protect\citeauthoryear{{Fattahi}, {Navarro}, {Starkenburg}, {Barber}
  \& {McConnachie}}{{Fattahi} et~al.}{2013}]{fattahi2013}
{Fattahi} A.,  {Navarro} J.~F.,  {Starkenburg} E.,  {Barber} C.~R.,
  {McConnachie} A.~W.,  2013, \mn@doi [\mnras] {10.1093/mnrasl/slt011}, \href
  {https://ui.adsabs.harvard.edu/abs/2013MNRAS.431L..73F} {431, L73}

\bibitem[\protect\citeauthoryear{{Faucher-Gigu{\`e}re}, {Lidz}, {Zaldarriaga}
  \& {Hernquist}}{{Faucher-Gigu{\`e}re} et~al.}{2009}]{faucher2009ionizing}
{Faucher-Gigu{\`e}re} C.-A.,  {Lidz} A.,  {Zaldarriaga} M.,   {Hernquist} L.,
  2009, \mn@doi [\apj] {10.1088/0004-637X/703/2/1416}, \href
  {http://adsabs.harvard.edu/abs/2009ApJ...703.1416F} {703, 1416}

\bibitem[\protect\citeauthoryear{{Ferrero}, {Abadi}, {Navarro}, {Sales}  \&
  {Gurovich}}{{Ferrero} et~al.}{2012}]{ferrero2012dwarfhalos}
{Ferrero} I.,  {Abadi} M.~G.,  {Navarro} J.~F.,  {Sales} L.~V.,   {Gurovich}
  S.,  2012, \mn@doi [\mnras] {10.1111/j.1365-2966.2012.21623.x}, \href
  {https://ui.adsabs.harvard.edu/abs/2012MNRAS.425.2817F} {425, 2817}

\bibitem[\protect\citeauthoryear{{Fitts} et~al.,}{{Fitts}
  et~al.}{2017}]{fitts2017field}
{Fitts} A.,  et~al., 2017, \mn@doi [\mnras] {10.1093/mnras/stx1757}, \href
  {https://ui.adsabs.harvard.edu/abs/2017MNRAS.471.3547F} {471, 3547}

\bibitem[\protect\citeauthoryear{{Gaia Collaboration} et~al.,}{{Gaia
  Collaboration} et~al.}{2018}]{helmi2018gaia}
{Gaia Collaboration} et~al., 2018, \mn@doi [\aap]
  {10.1051/0004-6361/201832698}, \href
  {https://ui.adsabs.harvard.edu/abs/2018A&A...616A..12G} {616, A12}

\bibitem[\protect\citeauthoryear{{Gao}, {White}, {Jenkins}, {Stoehr}  \&
  {Springel}}{{Gao} et~al.}{2004}]{gao2004SHpop}
{Gao} L.,  {White} S.~D.~M.,  {Jenkins} A.,  {Stoehr} F.,   {Springel} V.,
  2004, \mn@doi [\mnras] {10.1111/j.1365-2966.2004.08360.x}, \href
  {https://ui.adsabs.harvard.edu/abs/2004MNRAS.355..819G} {355, 819}

\bibitem[\protect\citeauthoryear{{Garrison-Kimmel}, {Boylan-Kolchin}, {Bullock}
   \& {Lee}}{{Garrison-Kimmel} et~al.}{2014}]{GK14elvis}
{Garrison-Kimmel} S.,  {Boylan-Kolchin} M.,  {Bullock} J.~S.,   {Lee} K.,
  2014, \mn@doi [\mnras] {10.1093/mnras/stt2377}, \href
  {https://ui.adsabs.harvard.edu/abs/2014MNRAS.438.2578G} {438, 2578}

\bibitem[\protect\citeauthoryear{{Garrison-Kimmel}, {Bullock}, {Boylan-Kolchin}
   \& {Bardwell}}{{Garrison-Kimmel} et~al.}{2017a}]{gk2017mstrscatter}
{Garrison-Kimmel} S.,  {Bullock} J.~S.,  {Boylan-Kolchin} M.,   {Bardwell} E.,
  2017a, \mn@doi [\mnras] {10.1093/mnras/stw2564}, \href
  {https://ui.adsabs.harvard.edu/abs/2017MNRAS.464.3108G} {464, 3108}

\bibitem[\protect\citeauthoryear{{Garrison-Kimmel} et~al.,}{{Garrison-Kimmel}
  et~al.}{2017b}]{notsolumpy}
{Garrison-Kimmel} S.,  et~al., 2017b, \mn@doi [\mnras] {10.1093/mnras/stx1710},
  \href {https://ui.adsabs.harvard.edu/abs/2017MNRAS.471.1709G} {471, 1709}

\bibitem[\protect\citeauthoryear{{Garrison-Kimmel} et~al.,}{{Garrison-Kimmel}
  et~al.}{2018}]{gk2018morph}
{Garrison-Kimmel} S.,  et~al., 2018, \mn@doi [\mnras] {10.1093/mnras/sty2513},
  \href {https://ui.adsabs.harvard.edu/abs/2018MNRAS.481.4133G} {481, 4133}

\bibitem[\protect\citeauthoryear{{Garrison-Kimmel} et~al.,}{{Garrison-Kimmel}
  et~al.}{2019}]{gk2019sfh}
{Garrison-Kimmel} S.,  et~al., 2019, \mn@doi [\mnras] {10.1093/mnras/stz1317},
  \href {https://ui.adsabs.harvard.edu/abs/2019MNRAS.tmp.1262G} {p.~1262}

\bibitem[\protect\citeauthoryear{{Giocoli}, {Tormen}  \& {van den
  Bosch}}{{Giocoli} et~al.}{2008}]{giocoli2008}
{Giocoli} C.,  {Tormen} G.,   {van den Bosch} F.~C.,  2008, \mn@doi [\mnras]
  {10.1111/j.1365-2966.2008.13182.x}, \href
  {http://adsabs.harvard.edu/abs/2008MNRAS.386.2135G} {386, 2135}

\bibitem[\protect\citeauthoryear{{Governato} et~al.,}{{Governato}
  et~al.}{2004}]{governato2004}
{Governato} F.,  et~al., 2004, \mn@doi [\apj] {10.1086/383516}, \href
  {https://ui.adsabs.harvard.edu/abs/2004ApJ...607..688G} {607, 688}

\bibitem[\protect\citeauthoryear{{Graus}, {Bullock}, {Kelley},
  {Boylan-Kolchin}, {Garrison-Kimmel}  \& {Qi}}{{Graus}
  et~al.}{2018}]{graus2018sats}
{Graus} A.~S.,  {Bullock} J.~S.,  {Kelley} T.,  {Boylan-Kolchin} M.,
  {Garrison-Kimmel} S.,   {Qi} Y.,  2018, preprint, \href
  {http://adsabs.harvard.edu/abs/2018arXiv180803654G} {} (\mn@eprint {arXiv}
  {1808.03654})

\bibitem[\protect\citeauthoryear{{Griffen}, {Ji}, {Dooley}, {G{\'o}mez},
  {Vogelsberger}, {O'Shea}  \& {Frebel}}{{Griffen}
  et~al.}{2016}]{Griffen16caterpillar}
{Griffen} B.~F.,  {Ji} A.~P.,  {Dooley} G.~A.,  {G{\'o}mez} F.~A.,
  {Vogelsberger} M.,  {O'Shea} B.~W.,   {Frebel} A.,  2016, \mn@doi [\apj]
  {10.3847/0004-637X/818/1/10}, \href
  {https://ui.adsabs.harvard.edu/abs/2016ApJ...818...10G} {818, 10}

\bibitem[\protect\citeauthoryear{{Guo}, {White}, {Li}  \&
  {Boylan-Kolchin}}{{Guo} et~al.}{2010}]{guo10smhm}
{Guo} Q.,  {White} S.,  {Li} C.,   {Boylan-Kolchin} M.,  2010, \mn@doi [\mnras]
  {10.1111/j.1365-2966.2010.16341.x}, \href
  {http://adsabs.harvard.edu/abs/2010MNRAS.404.1111G} {404, 1111}

\bibitem[\protect\citeauthoryear{{Guo} et~al.,}{{Guo} et~al.}{2011}]{guo2011}
{Guo} Q.,  et~al., 2011, \mn@doi [\mnras] {10.1111/j.1365-2966.2010.18114.x},
  \href {https://ui.adsabs.harvard.edu/abs/2011MNRAS.413..101G} {413, 101}

\bibitem[\protect\citeauthoryear{{Hahn} \& {Abel}}{{Hahn} \&
  {Abel}}{2011}]{hahn2011music}
{Hahn} O.,  {Abel} T.,  2011, \mn@doi [\mnras]
  {10.1111/j.1365-2966.2011.18820.x}, \href
  {https://ui.adsabs.harvard.edu/abs/2011MNRAS.415.2101H} {415, 2101}

\bibitem[\protect\citeauthoryear{{Hayashi}, {Navarro}, {Taylor}, {Stadel}  \&
  {Quinn}}{{Hayashi} et~al.}{2003}]{Hayashi2003}
{Hayashi} E.,  {Navarro} J.~F.,  {Taylor} J.~E.,  {Stadel} J.,   {Quinn} T.,
  2003, \mn@doi [\apj] {10.1086/345788}, \href
  {https://ui.adsabs.harvard.edu/abs/2003ApJ...584..541H} {584, 541}

\bibitem[\protect\citeauthoryear{{Hopkins}}{{Hopkins}}{2015}]{hopkins2015gizmo}
{Hopkins} P.~F.,  2015, \mn@doi [\mnras] {10.1093/mnras/stv195}, \href
  {https://ui.adsabs.harvard.edu/abs/2015MNRAS.450...53H} {450, 53}

\bibitem[\protect\citeauthoryear{{Hopkins}, {Narayanan}  \& {Murray}}{{Hopkins}
  et~al.}{2013}]{hopkins2013sf}
{Hopkins} P.~F.,  {Narayanan} D.,   {Murray} N.,  2013, \mn@doi [\mnras]
  {10.1093/mnras/stt723}, \href
  {https://ui.adsabs.harvard.edu/abs/2013MNRAS.432.2647H} {432, 2647}

\bibitem[\protect\citeauthoryear{{Hopkins} et~al.,}{{Hopkins}
  et~al.}{2018a}]{hopkins2018sne}
{Hopkins} P.~F.,  et~al., 2018a, \mn@doi [\mnras] {10.1093/mnras/sty674}, \href
  {https://ui.adsabs.harvard.edu/abs/2018MNRAS.477.1578H} {477, 1578}

\bibitem[\protect\citeauthoryear{{Hopkins} et~al.,}{{Hopkins}
  et~al.}{2018b}]{hopkins2018fire2}
{Hopkins} P.~F.,  et~al., 2018b, \mn@doi [\mnras] {10.1093/mnras/sty1690},
  \href {https://ui.adsabs.harvard.edu/abs/2018MNRAS.480..800H} {480, 800}

\bibitem[\protect\citeauthoryear{{Jenkins} et~al.,}{{Jenkins}
  et~al.}{1998}]{jenkins1998structure}
{Jenkins} A.,  et~al., 1998, \mn@doi [\apj] {10.1086/305615}, \href
  {https://ui.adsabs.harvard.edu/abs/1998ApJ...499...20J} {499, 20}

\bibitem[\protect\citeauthoryear{{Jethwa}, {Erkal}  \& {Belokurov}}{{Jethwa}
  et~al.}{2016}]{jethwa2016magDES}
{Jethwa} P.,  {Erkal} D.,   {Belokurov} V.,  2016, \mn@doi [\mnras]
  {10.1093/mnras/stw1343}, \href
  {https://ui.adsabs.harvard.edu/abs/2016MNRAS.461.2212J} {461, 2212}

\bibitem[\protect\citeauthoryear{{Kallivayalil}, {van der Marel}, {Besla},
  {Anderson}  \& {Alcock}}{{Kallivayalil} et~al.}{2013}]{kallivayalil2013}
{Kallivayalil} N.,  {van der Marel} R.~P.,  {Besla} G.,  {Anderson} J.,
  {Alcock} C.,  2013, \mn@doi [\apj] {10.1088/0004-637X/764/2/161}, \href
  {http://adsabs.harvard.edu/abs/2013ApJ...764..161K} {764, 161}

\bibitem[\protect\citeauthoryear{{Kallivayalil} et~al.,}{{Kallivayalil}
  et~al.}{2018}]{kallivayalil2018MCsats}
{Kallivayalil} N.,  et~al., 2018, \mn@doi [\apj] {10.3847/1538-4357/aadfee},
  \href {https://ui.adsabs.harvard.edu/abs/2018ApJ...867...19K} {867, 19}

\bibitem[\protect\citeauthoryear{{Katz} \& {White}}{{Katz} \&
  {White}}{1993}]{katzwhite1993}
{Katz} N.,  {White} S. D.~M.,  1993, \mn@doi [\apj] {10.1086/172935}, \href
  {https://ui.adsabs.harvard.edu/abs/1993ApJ...412..455K} {412, 455}

\bibitem[\protect\citeauthoryear{{Kelley}, {Bullock}, {Garrison-Kimmel},
  {Boylan-Kolchin}, {Pawlowski}  \& {Graus}}{{Kelley}
  et~al.}{2019}]{kelley18phatelvis}
{Kelley} T.,  {Bullock} J.~S.,  {Garrison-Kimmel} S.,  {Boylan-Kolchin} M.,
  {Pawlowski} M.~S.,   {Graus} A.~S.,  2019, \mn@doi [\mnras]
  {10.1093/mnras/stz1553}, \href
  {https://ui.adsabs.harvard.edu/abs/2019MNRAS.tmp.1496K} {p.~1496}

\bibitem[\protect\citeauthoryear{{Kim} \& {Jerjen}}{{Kim} \&
  {Jerjen}}{2015}]{kimJergen2015hor2}
{Kim} D.,  {Jerjen} H.,  2015, \mn@doi [\apj] {10.1088/2041-8205/808/2/L39},
  \href {https://ui.adsabs.harvard.edu/abs/2015ApJ...808L..39K} {808, L39}

\bibitem[\protect\citeauthoryear{{Kirby}, {Bullock}, {Boylan-Kolchin},
  {Kaplinghat}  \& {Cohen}}{{Kirby} et~al.}{2014}]{kirby2014dynamics}
{Kirby} E.~N.,  {Bullock} J.~S.,  {Boylan-Kolchin} M.,  {Kaplinghat} M.,
  {Cohen} J.~G.,  2014, \mn@doi [\mnras] {10.1093/mnras/stu025}, \href
  {https://ui.adsabs.harvard.edu/abs/2014MNRAS.439.1015K} {439, 1015}

\bibitem[\protect\citeauthoryear{{Klypin}, {Kravtsov}, {Valenzuela}  \&
  {Prada}}{{Klypin} et~al.}{1999}]{klypin1999missing}
{Klypin} A.,  {Kravtsov} A.~V.,  {Valenzuela} O.,   {Prada} F.,  1999, \mn@doi
  [\apj] {10.1086/307643}, \href
  {https://ui.adsabs.harvard.edu/abs/1999ApJ...522...82K} {522, 82}

\bibitem[\protect\citeauthoryear{{Koposov}, {Rix}  \& {Hogg}}{{Koposov}
  et~al.}{2010}]{koposov2010gd1}
{Koposov} S.~E.,  {Rix} H.-W.,   {Hogg} D.~W.,  2010, \mn@doi [\apj]
  {10.1088/0004-637X/712/1/260}, \href
  {http://adsabs.harvard.edu/abs/2010ApJ...712..260K} {712, 260}

\bibitem[\protect\citeauthoryear{{Koposov}, {Belokurov}, {Torrealba}  \&
  {Evans}}{{Koposov} et~al.}{2015}]{koposov2015DES}
{Koposov} S.~E.,  {Belokurov} V.,  {Torrealba} G.,   {Evans} N.~W.,  2015,
  \mn@doi [\apj] {10.1088/0004-637X/805/2/130}, \href
  {https://ui.adsabs.harvard.edu/abs/2015ApJ...805..130K} {805, 130}

\bibitem[\protect\citeauthoryear{{Koposov} et~al.,}{{Koposov}
  et~al.}{2018}]{koposov2018hyi}
{Koposov} S.~E.,  et~al., 2018, \mn@doi [\mnras] {10.1093/mnras/sty1772}, \href
  {https://ui.adsabs.harvard.edu/abs/2018MNRAS.479.5343K} {479, 5343}

\bibitem[\protect\citeauthoryear{{Kravtsov}, {Berlind}, {Wechsler}, {Klypin},
  {Gottl{\"o}ber}, {Allgood}  \& {Primack}}{{Kravtsov}
  et~al.}{2004}]{kravtsov2004halo}
{Kravtsov} A.~V.,  {Berlind} A.~A.,  {Wechsler} R.~H.,  {Klypin} A.~A.,
  {Gottl{\"o}ber} S.,  {Allgood} B.~o.,   {Primack} J.~R.,  2004, \mn@doi
  [\apj] {10.1086/420959}, \href
  {https://ui.adsabs.harvard.edu/abs/2004ApJ...609...35K} {609, 35}

\bibitem[\protect\citeauthoryear{{Kroupa}}{{Kroupa}}{2001}]{kroupa2001imf}
{Kroupa} P.,  2001, \mn@doi [\mnras] {10.1046/j.1365-8711.2001.04022.x}, \href
  {https://ui.adsabs.harvard.edu/abs/2001MNRAS.322..231K} {322, 231}

\bibitem[\protect\citeauthoryear{{Krumholz} \& {Gnedin}}{{Krumholz} \&
  {Gnedin}}{2011}]{Krumholz2011}
{Krumholz} M.~R.,  {Gnedin} N.~Y.,  2011, \mn@doi [\apj]
  {10.1088/0004-637X/729/1/36}, \href
  {http://adsabs.harvard.edu/abs/2011ApJ...729...36K} {729, 36}

\bibitem[\protect\citeauthoryear{{Laevens} et~al.,}{{Laevens}
  et~al.}{2015}]{laevens2015MWsats}
{Laevens} B. P.~M.,  et~al., 2015, \mn@doi [\apj] {10.1088/0004-637X/813/1/44},
  \href {https://ui.adsabs.harvard.edu/abs/2015ApJ...813...44L} {813, 44}

\bibitem[\protect\citeauthoryear{{Leitherer} et~al.,}{{Leitherer}
  et~al.}{1999}]{leitherer1999starburst}
{Leitherer} C.,  et~al., 1999, \mn@doi [The Astrophysical Journal Supplement
  Series] {10.1086/313233}, \href
  {https://ui.adsabs.harvard.edu/abs/1999ApJS..123....3L} {123, 3}

\bibitem[\protect\citeauthoryear{{Lynden-Bell} \& {Lynden-Bell}}{{Lynden-Bell}
  \& {Lynden-Bell}}{1995}]{lyndenbell1995}
{Lynden-Bell} D.,  {Lynden-Bell} R.~M.,  1995, \mn@doi [\mnras]
  {10.1093/mnras/275.2.429}, \href
  {https://ui.adsabs.harvard.edu/abs/1995MNRAS.275..429L} {275, 429}

\bibitem[\protect\citeauthoryear{{Ma}, {Hopkins}, {Faucher-Gigu{\`e}re},
  {Zolman}, {Muratov}, {Kere{\v{s}}}  \& {Quataert}}{{Ma}
  et~al.}{2016}]{ma2016metallicity}
{Ma} X.,  {Hopkins} P.~F.,  {Faucher-Gigu{\`e}re} C.-A.,  {Zolman} N.,
  {Muratov} A.~L.,  {Kere{\v{s}}} D.,   {Quataert} E.,  2016, \mn@doi [\mnras]
  {10.1093/mnras/stv2659}, \href
  {https://ui.adsabs.harvard.edu/abs/2016MNRAS.456.2140M} {456, 2140}

\bibitem[\protect\citeauthoryear{{McConnachie}}{{McConnachie}}{2012}]{mcconn2012DG}
{McConnachie} A.~W.,  2012, \mn@doi [\aj] {10.1088/0004-6256/144/1/4}, \href
  {https://ui.adsabs.harvard.edu/abs/2012AJ....144....4M} {144, 4}

\bibitem[\protect\citeauthoryear{{Metcalf} \& {Zhao}}{{Metcalf} \&
  {Zhao}}{2002}]{metcalfzhou02fluxratios}
{Metcalf} R.~B.,  {Zhao} H.,  2002, \mn@doi [\apjl] {10.1086/339798}, \href
  {http://adsabs.harvard.edu/abs/2002ApJ...567L...5M} {567, L5}

\bibitem[\protect\citeauthoryear{{Moore}, {Ghigna}, {Governato}, {Lake},
  {Quinn}, {Stadel}  \& {Tozzi}}{{Moore} et~al.}{1999}]{moore1999}
{Moore} B.,  {Ghigna} S.,  {Governato} F.,  {Lake} G.,  {Quinn} T.,  {Stadel}
  J.,   {Tozzi} P.,  1999, \mn@doi [\apj] {10.1086/312287}, \href
  {https://ui.adsabs.harvard.edu/abs/1999ApJ...524L..19M} {524, L19}

\bibitem[\protect\citeauthoryear{{Moster}, {Naab}  \& {White}}{{Moster}
  et~al.}{2013}]{moster2013halos}
{Moster} B.~P.,  {Naab} T.,   {White} S.~D.~M.,  2013, \mn@doi [\mnras]
  {10.1093/mnras/sts261}, \href
  {http://adsabs.harvard.edu/abs/2013MNRAS.428.3121M} {428, 3121}

\bibitem[\protect\citeauthoryear{{Navarro}, {Frenk}  \& {White}}{{Navarro}
  et~al.}{1996}]{navarro1996CDMhalos}
{Navarro} J.~F.,  {Frenk} C.~S.,   {White} S. D.~M.,  1996, \mn@doi [\apj]
  {10.1086/177173}, \href
  {https://ui.adsabs.harvard.edu/abs/1996ApJ...462..563N} {462, 563}

\bibitem[\protect\citeauthoryear{{Nierenberg}, {Treu}, {Wright}, {Fassnacht}
  \& {Auger}}{{Nierenberg} et~al.}{2014}]{nierenberg2014detect}
{Nierenberg} A.~M.,  {Treu} T.,  {Wright} S.~A.,  {Fassnacht} C.~D.,   {Auger}
  M.~W.,  2014, \mn@doi [\mnras] {10.1093/mnras/stu862}, \href
  {http://adsabs.harvard.edu/abs/2014MNRAS.442.2434N} {442, 2434}

\bibitem[\protect\citeauthoryear{{O{\~n}orbe}, {Boylan-Kolchin}, {Bullock},
  {Hopkins}, {Kere{\v{s}}}, {Faucher-Gigu{\`e}re}, {Quataert}  \&
  {Murray}}{{O{\~n}orbe} et~al.}{2015}]{onorbe2015fire}
{O{\~n}orbe} J.,  {Boylan-Kolchin} M.,  {Bullock} J.~S.,  {Hopkins} P.~F.,
  {Kere{\v{s}}} D.,  {Faucher-Gigu{\`e}re} C.-A.,  {Quataert} E.,   {Murray}
  N.,  2015, \mn@doi [\mnras] {10.1093/mnras/stv2072}, \href
  {https://ui.adsabs.harvard.edu/abs/2015MNRAS.454.2092O} {454, 2092}

\bibitem[\protect\citeauthoryear{{Okamoto} \& {Frenk}}{{Okamoto} \&
  {Frenk}}{2009}]{okamotoFrenk2009}
{Okamoto} T.,  {Frenk} C.~S.,  2009, \mn@doi [\mnras]
  {10.1111/j.1745-3933.2009.00748.x}, \href
  {https://ui.adsabs.harvard.edu/abs/2009MNRAS.399L.174O} {399, L174}

\bibitem[\protect\citeauthoryear{{Orr} et~al.,}{{Orr} et~al.}{2018}]{orr2018ks}
{Orr} M.~E.,  et~al., 2018, \mn@doi [\mnras] {10.1093/mnras/sty1241}, \href
  {https://ui.adsabs.harvard.edu/abs/2018MNRAS.478.3653O} {478, 3653}

\bibitem[\protect\citeauthoryear{{Pardy} et~al.,}{{Pardy}
  et~al.}{2019}]{pardy2019aurigaLMC}
{Pardy} S.~A.,  et~al., 2019, arXiv e-prints, \href
  {http://adsabs.harvard.edu/abs/2019arXiv190401028P} {}

\bibitem[\protect\citeauthoryear{{Pe{\~n}arrubia}, {Benson}, {Walker},
  {Gilmore}, {McConnachie}  \& {Mayer}}{{Pe{\~n}arrubia}
  et~al.}{2010}]{penarrubia2010cores}
{Pe{\~n}arrubia} J.,  {Benson} A.~J.,  {Walker} M.~G.,  {Gilmore} G.,
  {McConnachie} A.~W.,   {Mayer} L.,  2010, \mn@doi [\mnras]
  {10.1111/j.1365-2966.2010.16762.x}, \href
  {https://ui.adsabs.harvard.edu/abs/2010MNRAS.406.1290P} {406, 1290}

\bibitem[\protect\citeauthoryear{{Pontzen} \& {Governato}}{{Pontzen} \&
  {Governato}}{2012}]{pontGov2012SNe}
{Pontzen} A.,  {Governato} F.,  2012, \mn@doi [\mnras]
  {10.1111/j.1365-2966.2012.20571.x}, \href
  {https://ui.adsabs.harvard.edu/abs/2012MNRAS.421.3464P} {421, 3464}

\bibitem[\protect\citeauthoryear{{Power}, {Navarro}, {Jenkins}, {Frenk},
  {White}, {Springel}, {Stadel}  \& {Quinn}}{{Power} et~al.}{2003}]{power03dm}
{Power} C.,  {Navarro} J.~F.,  {Jenkins} A.,  {Frenk} C.~S.,  {White} S.~D.~M.,
   {Springel} V.,  {Stadel} J.,   {Quinn} T.,  2003, \mn@doi [\mnras]
  {10.1046/j.1365-8711.2003.05925.x}, \href
  {http://adsabs.harvard.edu/abs/2003MNRAS.338...14P} {338, 14}

\bibitem[\protect\citeauthoryear{{Press} \& {Schechter}}{{Press} \&
  {Schechter}}{1974}]{pressSchechter1974}
{Press} W.~H.,  {Schechter} P.,  1974, \mn@doi [\apj] {10.1086/152650}, \href
  {https://ui.adsabs.harvard.edu/abs/1974ApJ...187..425P} {187, 425}

\bibitem[\protect\citeauthoryear{{Price} \& {Monaghan}}{{Price} \&
  {Monaghan}}{2007}]{price2007}
{Price} D.~J.,  {Monaghan} J.~J.,  2007, \mn@doi [\mnras]
  {10.1111/j.1365-2966.2006.11241.x}, \href
  {https://ui.adsabs.harvard.edu/abs/2007MNRAS.374.1347P} {374, 1347}

\bibitem[\protect\citeauthoryear{{Read}, {Wilkinson}, {Evans}, {Gilmore}  \&
  {Kleyna}}{{Read} et~al.}{2006}]{Read2006tides}
{Read} J.~I.,  {Wilkinson} M.~I.,  {Evans} N.~W.,  {Gilmore} G.,   {Kleyna}
  J.~T.,  2006, \mn@doi [\mnras] {10.1111/j.1365-2966.2005.09959.x}, \href
  {https://ui.adsabs.harvard.edu/abs/2006MNRAS.367..387R} {367, 387}

\bibitem[\protect\citeauthoryear{{Sales}, {Navarro}, {Abadi}  \&
  {Steinmetz}}{{Sales} et~al.}{2007}]{sales2007simSats}
{Sales} L.~V.,  {Navarro} J.~F.,  {Abadi} M.~G.,   {Steinmetz} M.,  2007,
  \mn@doi [\mnras] {10.1111/j.1365-2966.2007.12024.x}, \href
  {https://ui.adsabs.harvard.edu/abs/2007MNRAS.379.1464S} {379, 1464}

\bibitem[\protect\citeauthoryear{{Sales}, {Navarro}, {Cooper}, {White}, {Frenk}
   \& {Helmi}}{{Sales} et~al.}{2011}]{sales2011MagGal}
{Sales} L.~V.,  {Navarro} J.~F.,  {Cooper} A.~P.,  {White} S. D.~M.,  {Frenk}
  C.~S.,   {Helmi} A.,  2011, \mn@doi [\mnras]
  {10.1111/j.1365-2966.2011.19514.x}, \href
  {https://ui.adsabs.harvard.edu/abs/2011MNRAS.418..648S} {418, 648}

\bibitem[\protect\citeauthoryear{{Sales}, {Wang}, {White}  \&
  {Navarro}}{{Sales} et~al.}{2013}]{sales2013dwfsats}
{Sales} L.~V.,  {Wang} W.,  {White} S. D.~M.,   {Navarro} J.~F.,  2013, \mn@doi
  [\mnras] {10.1093/mnras/sts054}, \href
  {https://ui.adsabs.harvard.edu/abs/2013MNRAS.428..573S} {428, 573}

\bibitem[\protect\citeauthoryear{{Sales}, {Navarro}, {Kallivayalil}  \&
  {Frenk}}{{Sales} et~al.}{2017}]{sales2017truMCsats}
{Sales} L.~V.,  {Navarro} J.~F.,  {Kallivayalil} N.,   {Frenk} C.~S.,  2017,
  \mn@doi [\mnras] {10.1093/mnras/stw2816}, \href
  {https://ui.adsabs.harvard.edu/abs/2017MNRAS.465.1879S} {465, 1879}

\bibitem[\protect\citeauthoryear{{Samuel} et~al.,}{{Samuel}
  et~al.}{2019}]{samuel2019radial}
{Samuel} J.,  et~al., 2019, arXiv e-prints, \href
  {https://ui.adsabs.harvard.edu/abs/2019arXiv190411508S} {p. arXiv:1904.11508}

\bibitem[\protect\citeauthoryear{{Sawala} et~al.,}{{Sawala}
  et~al.}{2016}]{sawala2016}
{Sawala} T.,  et~al., 2016, \mn@doi [\mnras] {10.1093/mnras/stw145}, \href
  {http://adsabs.harvard.edu/abs/2016MNRAS.457.1931S} {457, 1931}

\bibitem[\protect\citeauthoryear{{Sawala}, {Pihajoki}, {Johansson}, {Frenk},
  {Navarro}, {Oman}  \& {White}}{{Sawala} et~al.}{2017}]{sawala2017}
{Sawala} T.,  {Pihajoki} P.,  {Johansson} P.~H.,  {Frenk} C.~S.,  {Navarro}
  J.~F.,  {Oman} K.~A.,   {White} S. D.~M.,  2017, \mn@doi [\mnras]
  {10.1093/mnras/stx360}, \href
  {https://ui.adsabs.harvard.edu/abs/2017MNRAS.467.4383S} {467, 4383}

\bibitem[\protect\citeauthoryear{{Shao}, {Cautun}, {Frenk}, {Grand },
  {G{\'o}mez}, {Marinacci}  \& {Simpson}}{{Shao}
  et~al.}{2018a}]{shao2018satacc}
{Shao} S.,  {Cautun} M.,  {Frenk} C.~S.,  {Grand } R. J.~J.,  {G{\'o}mez}
  F.~A.,  {Marinacci} F.,   {Simpson} C.~M.,  2018a, \mn@doi [\mnras]
  {10.1093/mnras/sty343}, \href
  {https://ui.adsabs.harvard.edu/abs/2018MNRAS.476.1796S} {476, 1796}

\bibitem[\protect\citeauthoryear{{Shao}, {Cautun}, {Deason}, {Frenk}  \&
  {Theuns}}{{Shao} et~al.}{2018b}]{shao2018lmceagle}
{Shao} S.,  {Cautun} M.,  {Deason} A.~J.,  {Frenk} C.~S.,   {Theuns} T.,
  2018b, \mn@doi [\mnras] {10.1093/mnras/sty1470}, \href
  {https://ui.adsabs.harvard.edu/abs/2018MNRAS.479..284S} {479, 284}

\bibitem[\protect\citeauthoryear{{Simon}}{{Simon}}{2018}]{simon2018GAIA}
{Simon} J.~D.,  2018, \mn@doi [\apj] {10.3847/1538-4357/aacdfb}, \href
  {http://adsabs.harvard.edu/abs/2018ApJ...863...89S} {863, 89}

\bibitem[\protect\citeauthoryear{{Simon}}{{Simon}}{2019}]{simon2019UFDs}
{Simon} J.~D.,  2019, arXiv e-prints, \href
  {https://ui.adsabs.harvard.edu/\#abs/2019arXiv190105465S} {p.
  arXiv:1901.05465}

\bibitem[\protect\citeauthoryear{{Simpson}, {Grand}, {G{\'o}mez}, {Marinacci},
  {Pakmor}, {Springel}, {Campbell}  \& {Frenk}}{{Simpson}
  et~al.}{2018}]{simpson2018auriga}
{Simpson} C.~M.,  {Grand} R. J.~J.,  {G{\'o}mez} F.~A.,  {Marinacci} F.,
  {Pakmor} R.,  {Springel} V.,  {Campbell} D. J.~R.,   {Frenk} C.~S.,  2018,
  \mn@doi [\mnras] {10.1093/mnras/sty774}, \href
  {https://ui.adsabs.harvard.edu/abs/2018MNRAS.478..548S} {478, 548}

\bibitem[\protect\citeauthoryear{{Somerville}}{{Somerville}}{2002}]{somerville02photo}
{Somerville} R.~S.,  2002, \mn@doi [\apj] {10.1086/341444}, \href
  {https://ui.adsabs.harvard.edu/abs/2002ApJ...572L..23S} {572, L23}

\bibitem[\protect\citeauthoryear{{Springel}}{{Springel}}{2005}]{springel2005gadget2}
{Springel} V.,  2005, \mn@doi [\mnras] {10.1111/j.1365-2966.2005.09655.x},
  \href {https://ui.adsabs.harvard.edu/abs/2005MNRAS.364.1105S} {364, 1105}

\bibitem[\protect\citeauthoryear{{Springel} et~al.,}{{Springel}
  et~al.}{2005}]{springel2005millenium}
{Springel} V.,  et~al., 2005, \mn@doi [\nat] {10.1038/nature03597}, \href
  {https://ui.adsabs.harvard.edu/abs/2005Natur.435..629S} {435, 629}

\bibitem[\protect\citeauthoryear{{Springel} et~al.,}{{Springel}
  et~al.}{2008}]{Springel08Aq}
{Springel} V.,  et~al., 2008, \mn@doi [\mnras]
  {10.1111/j.1365-2966.2008.14066.x}, \href
  {https://ui.adsabs.harvard.edu/abs/2008MNRAS.391.1685S} {391, 1685}

\bibitem[\protect\citeauthoryear{{Stierwalt}, {Besla}, {Patton}, {Johnson},
  {Kallivayalil}, {Putman}, {Privon}  \& {Ross}}{{Stierwalt}
  et~al.}{2015}]{stierwalt2015}
{Stierwalt} S.,  {Besla} G.,  {Patton} D.,  {Johnson} K.,  {Kallivayalil} N.,
  {Putman} M.,  {Privon} G.,   {Ross} G.,  2015, \mn@doi [\apj]
  {10.1088/0004-637X/805/1/2}, \href
  {http://adsabs.harvard.edu/abs/2015ApJ...805....2S} {805, 2}

\bibitem[\protect\citeauthoryear{{Strigari}, {Frenk}  \& {White}}{{Strigari}
  et~al.}{2010}]{strigari2010kinematics}
{Strigari} L.~E.,  {Frenk} C.~S.,   {White} S.~D.~M.,  2010, \mn@doi [\mnras]
  {10.1111/j.1365-2966.2010.17287.x}, \href
  {http://adsabs.harvard.edu/abs/2010MNRAS.408.2364S} {408, 2364}

\bibitem[\protect\citeauthoryear{{Su}, {Hopkins}, {Hayward},
  {Faucher-Gigu{\`e}re}, {Kere{\v{s}}}, {Ma}  \& {Robles}}{{Su}
  et~al.}{2017}]{su2017feedback}
{Su} K.-Y.,  {Hopkins} P.~F.,  {Hayward} C.~C.,  {Faucher-Gigu{\`e}re} C.-A.,
  {Kere{\v{s}}} D.,  {Ma} X.,   {Robles} V.~H.,  2017, \mn@doi [\mnras]
  {10.1093/mnras/stx1463}, \href
  {https://ui.adsabs.harvard.edu/abs/2017MNRAS.471..144S} {471, 144}

\bibitem[\protect\citeauthoryear{{Tegmark}, {Silk}, {Rees}, {Blanchard}, {Abel}
   \& {Palla}}{{Tegmark} et~al.}{1997}]{tegmark1997}
{Tegmark} M.,  {Silk} J.,  {Rees} M.~J.,  {Blanchard} A.,  {Abel} T.,   {Palla}
  F.,  1997, \mn@doi [\apj] {10.1086/303434}, \href
  {https://ui.adsabs.harvard.edu/abs/1997ApJ...474....1T} {474, 1}

\bibitem[\protect\citeauthoryear{{Tollerud}, {Boylan-Kolchin}  \&
  {Bullock}}{{Tollerud} et~al.}{2014}]{tollerud2014tbtf}
{Tollerud} E.~J.,  {Boylan-Kolchin} M.,   {Bullock} J.~S.,  2014, \mn@doi
  [\mnras] {10.1093/mnras/stu474}, \href
  {https://ui.adsabs.harvard.edu/abs/2014MNRAS.440.3511T} {440, 3511}

\bibitem[\protect\citeauthoryear{{Tollet} et~al.,}{{Tollet}
  et~al.}{2016}]{tollet2016nihaocores}
{Tollet} E.,  et~al., 2016, \mn@doi [\mnras] {10.1093/mnras/stv2856}, \href
  {https://ui.adsabs.harvard.edu/abs/2016MNRAS.456.3542T} {456, 3542}

\bibitem[\protect\citeauthoryear{{Torrealba} et~al.,}{{Torrealba}
  et~al.}{2016}]{torrealba2016Aq2}
{Torrealba} G.,  et~al., 2016, \mn@doi [\mnras] {10.1093/mnras/stw2051}, \href
  {https://ui.adsabs.harvard.edu/abs/2016MNRAS.463..712T} {463, 712}

\bibitem[\protect\citeauthoryear{{Torrealba} et~al.,}{{Torrealba}
  et~al.}{2018}]{torrealba2018antlia2}
{Torrealba} G.,  et~al., 2018, arXiv e-prints, \href
  {https://ui.adsabs.harvard.edu/abs/2018arXiv181104082T} {p. arXiv:1811.04082}

\bibitem[\protect\citeauthoryear{{Wang}, {Dutton}, {Stinson}, {Macci{\`o}},
  {Penzo}, {Kang}, {Keller}  \& {Wadsley}}{{Wang} et~al.}{2015}]{wang2015nihao}
{Wang} L.,  {Dutton} A.~A.,  {Stinson} G.~S.,  {Macci{\`o}} A.~V.,  {Penzo} C.,
   {Kang} X.,  {Keller} B.~W.,   {Wadsley} J.,  2015, \mn@doi [\mnras]
  {10.1093/mnras/stv1937}, \href
  {https://ui.adsabs.harvard.edu/abs/2015MNRAS.454...83W} {454, 83}

\bibitem[\protect\citeauthoryear{{Wetzel} \& {Nagai}}{{Wetzel} \&
  {Nagai}}{2015}]{wetzel2015accretion}
{Wetzel} A.~R.,  {Nagai} D.,  2015, \mn@doi [\apj]
  {10.1088/0004-637X/808/1/40}, \href
  {https://ui.adsabs.harvard.edu/abs/2015ApJ...808...40W} {808, 40}

\bibitem[\protect\citeauthoryear{{Wetzel}, {Deason}  \&
  {Garrison-Kimmel}}{{Wetzel} et~al.}{2015}]{wetzelDeasonGK2015}
{Wetzel} A.~R.,  {Deason} A.~J.,   {Garrison-Kimmel} S.,  2015, \mn@doi [\apj]
  {10.1088/0004-637X/807/1/49}, \href
  {https://ui.adsabs.harvard.edu/abs/2015ApJ...807...49W} {807, 49}

\bibitem[\protect\citeauthoryear{{Wetzel}, {Hopkins}, {Kim},
  {Faucher-Gigu{\`e}re}, {Kere{\v{s}}}  \& {Quataert}}{{Wetzel}
  et~al.}{2016}]{wetzel2016LATTE}
{Wetzel} A.~R.,  {Hopkins} P.~F.,  {Kim} J.-h.,  {Faucher-Gigu{\`e}re} C.-A.,
  {Kere{\v{s}}} D.,   {Quataert} E.,  2016, \mn@doi [\apj]
  {10.3847/2041-8205/827/2/L23}, \href
  {https://ui.adsabs.harvard.edu/abs/2016ApJ...827L..23W} {827, L23}

\bibitem[\protect\citeauthoryear{{Wheeler}, {O{\~n}orbe}, {Bullock},
  {Boylan-Kolchin}, {Elbert}, {Garrison-Kimmel}, {Hopkins}  \&
  {Kere{\v{s}}}}{{Wheeler} et~al.}{2015}]{wheeler2015}
{Wheeler} C.,  {O{\~n}orbe} J.,  {Bullock} J.~S.,  {Boylan-Kolchin} M.,
  {Elbert} O.~D.,  {Garrison-Kimmel} S.,  {Hopkins} P.~F.,   {Kere{\v{s}}} D.,
  2015, \mn@doi [\mnras] {10.1093/mnras/stv1691}, \href
  {https://ui.adsabs.harvard.edu/abs/2015MNRAS.453.1305W} {453, 1305}

\bibitem[\protect\citeauthoryear{{Wheeler} et~al.,}{{Wheeler}
  et~al.}{2018}]{wheeler2018}
{Wheeler} C.,  et~al., 2018, arXiv e-prints, \href
  {https://ui.adsabs.harvard.edu/abs/2018arXiv181202749W} {p. arXiv:1812.02749}

\bibitem[\protect\citeauthoryear{{White} \& {Rees}}{{White} \&
  {Rees}}{1978}]{white1978core}
{White} S.~D.~M.,  {Rees} M.~J.,  1978, \mn@doi [\mnras]
  {10.1093/mnras/183.3.341}, \href
  {https://ui.adsabs.harvard.edu/abs/1978MNRAS.183..341W} {183, 341}

\bibitem[\protect\citeauthoryear{{Wolf}, {Martinez}, {Bullock}, {Kaplinghat},
  {Geha}, {Mu{\~n}oz}, {Simon}  \& {Avedo}}{{Wolf}
  et~al.}{2010}]{wolf2010masses}
{Wolf} J.,  {Martinez} G.~D.,  {Bullock} J.~S.,  {Kaplinghat} M.,  {Geha} M.,
  {Mu{\~n}oz} R.~R.,  {Simon} J.~D.,   {Avedo} F.~F.,  2010, \mn@doi [\mnras]
  {10.1111/j.1365-2966.2010.16753.x}, \href
  {http://adsabs.harvard.edu/abs/2010MNRAS.406.1220W} {406, 1220}

\bibitem[\protect\citeauthoryear{{Wright} et~al.,}{{Wright}
  et~al.}{2017}]{wright2017GAMA}
{Wright} A.~H.,  et~al., 2017, \mn@doi [\mnras] {10.1093/mnras/stx1149}, \href
  {https://ui.adsabs.harvard.edu/abs/2017MNRAS.470..283W} {470, 283}

\bibitem[\protect\citeauthoryear{{Yang}, {Mo}  \& {van den Bosch}}{{Yang}
  et~al.}{2003}]{yang03galform}
{Yang} X.,  {Mo} H.~J.,   {van den Bosch} F.~C.,  2003, \mn@doi [\mnras]
  {10.1046/j.1365-8711.2003.06254.x}, \href
  {http://adsabs.harvard.edu/abs/2003MNRAS.339.1057Y} {339, 1057}

\bibitem[\protect\citeauthoryear{{Yang}, {Mo}, {Zhang}  \& {van den
  Bosch}}{{Yang} et~al.}{2011}]{yang2011}
{Yang} X.,  {Mo} H.~J.,  {Zhang} Y.,   {van den Bosch} F.~C.,  2011, \mn@doi
  [\apj] {10.1088/0004-637X/741/1/13}, \href
  {http://adsabs.harvard.edu/abs/2011ApJ...741...13Y} {741, 13}

\bibitem[\protect\citeauthoryear{{Yoon}, {Johnston}  \& {Hogg}}{{Yoon}
  et~al.}{2011}]{yoon2011cold}
{Yoon} J.~H.,  {Johnston} K.~V.,   {Hogg} D.~W.,  2011, \mn@doi [\apj]
  {10.1088/0004-637X/731/1/58}, \href
  {http://adsabs.harvard.edu/abs/2011ApJ...731...58Y} {731, 58}

\bibitem[\protect\citeauthoryear{{van der Marel} \& {Kallivayalil}}{{van der
  Marel} \& {Kallivayalil}}{2014}]{vanderMarel2014LMC}
{van der Marel} R.~P.,  {Kallivayalil} N.,  2014, \mn@doi [\apj]
  {10.1088/0004-637X/781/2/121}, \href
  {http://adsabs.harvard.edu/abs/2014ApJ...781..121V} {781, 121}

\makeatother
\end{thebibliography}

\appendix
\section{Additional Information}

\begin{figure}
    \centering
    \includegraphics[width=\columnwidth]{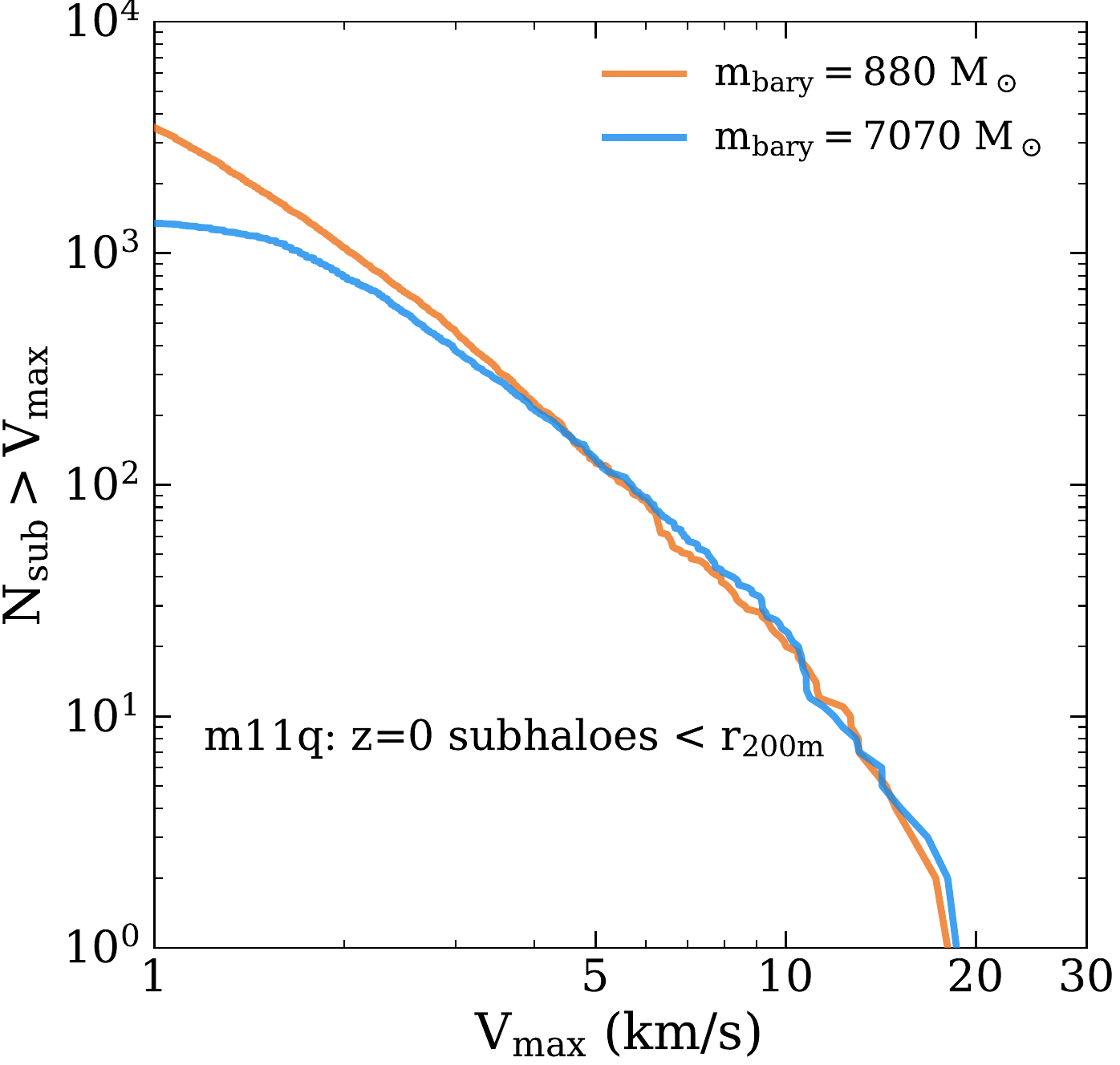}
    \caption{Resolution convergence of the $z=0$ subhalo population of \mqs at two resolutions: $m_\text{bary}$ = 7070 \msuns (blue) and $m_\text{bary}$ = 880 \msun (orange). The populations diverge below the typical value ($\sim$4 km/s) used for our minimum subhalo \vmaxs cutoff in Section \ref{sec:SubSuppression}, meaning the subhalo populations used in that analysis are well converged.}
    \label{fig:convergence}
\end{figure}

\begin{figure}
    \centering
    \includegraphics[width=\columnwidth]{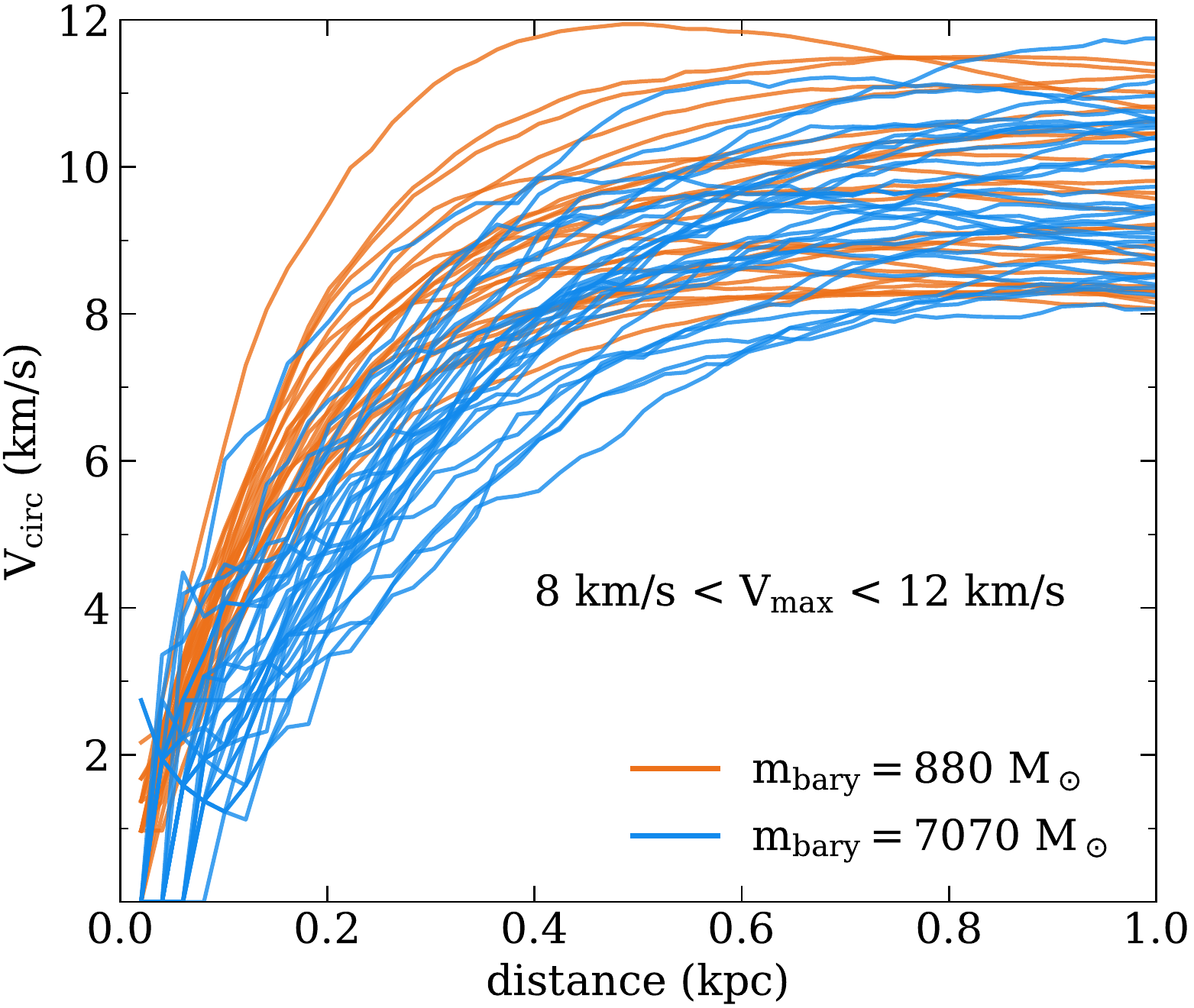}
    \caption{Circular velocity profiles (\vcircs = $\sqrt{GM(<r)/r}$) of subhalos with 8 km/s < \vmaxs < 12 km/s in \texttt{m11q} in high resolution ($m_\text{bary}$ = 880 \msun, orange) and the lowest resolution included herein ($m_\text{bary}$ = 7070 \msun, blue). The lower resolution run shows a systematic underdensity until convergence is reached at $r \sim 0.8$ kpc. Due to the integrated nature of these quantities, convergence in \vmaxs occurs at much higher radii ($\sim$ 800 pc) than the nominal softening lengths ($\epsilon_\text{DM} \sim 20 - 40$ pc). Distances at which ultrafaint circular velocities are measured are well below this convergence radius.}
    \label{fig:vcirc_restest}
\end{figure}

\begin{figure}
    \centering
    \includegraphics[width=\columnwidth]{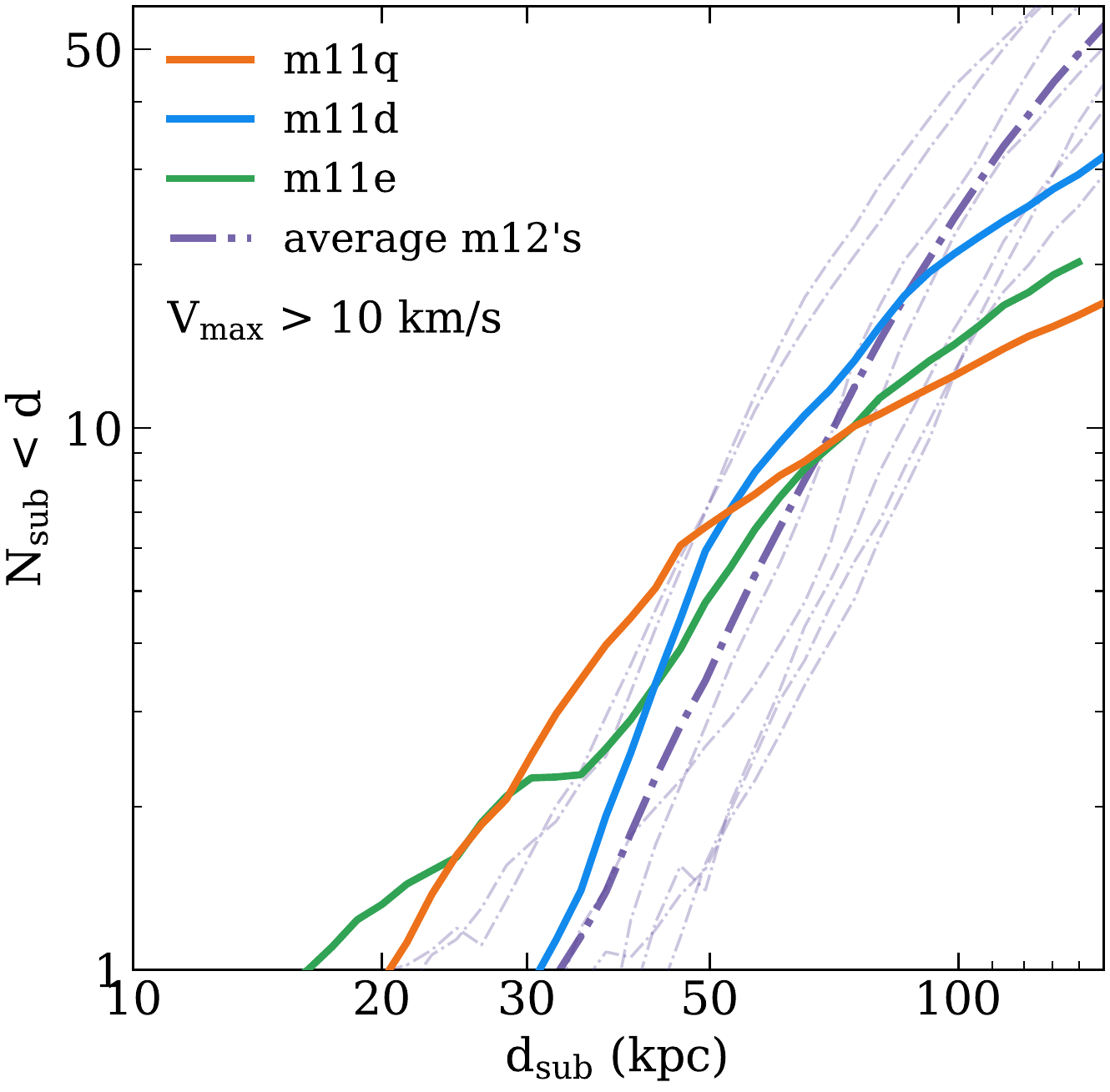}
    \caption{The radial distribution of subhalos with \vmaxs > 10 km/s around three LMC-mass hosts in FIRE (\texttt{m11q} - orange; \texttt{m11d - blue}; \texttt{m11e} - green) and averaged over seven MW-mass hosts in FIRE (purple dot-dash line), with all counts averaged over $\sim$1.3 Gyr. We find that two of the three LMC-mass halos host substructure at significantly smaller distances than the MW-mass halos, with all hosting more total subhalos than the average MW-mass halo out to $\sim$60 kpc. The existence of subhalos at closer distances to the central galaxy for LMC-mass hosts strengthens the plausibility of a detectable subhalo-galaxy interaction for hosts of this scale.}
    \label{fig:subprof} 
\end{figure}

Here we include additional information relevant to the discussions above. We test the effects of resolution in our simulations by comparing the subhalos of \texttt{m11q} run at the `high' and `normal' resolutions as described in sub-section \ref{ssec:LMCsample}. We examine the subhalo \vmaxs functions as described in sub-section \ref{ssec:resconv} in Figure \ref{fig:convergence} and the circular velocity profiles of subhalos with 8 km/s < \vmaxs < 12 km/s as described in sub-section \ref{ssec:dmcontent} in Figure \ref{fig:vcirc_restest}. We examine the radial distribution of subhalos in LMC-mass hosts and MW-mass hosts as described in sub-section \ref{ssec:subhalo_obs} in Figure \ref{fig:subprof}. We include our observational data set as described in sub-section~\ref{ssec:gaia} in Table \ref{tab:obs}, including stellar mass, galactocentric coordinates and radial velocities, distance to the LMC, as well as the current status of association to the LMC system.

{}

\newpage
\begin{table*}
\centering
\begin{tabular*}{\textwidth}{ @{\extracolsep{\fill}} l | c c c c c c | c c }
Name 		& \mstrs 		& $l$ & $b$ & $D_\text{MW}$ & $D_\text{LMC}$ & $V_\text{rad}$ (MW) & possible & confirmed		\\
 & (\msun) & (deg$^\circ$) & (deg$^\circ$) & (kpc) & (kpc)  & (km s$^{-1}$) & & \\
\hline

Antlia 2     & 4.30e5$^{\star, 6}$  & 264.9         & 11.2          & 124.2      & 145.4     & 49.6             &           & \\
Bootes I      & 2.90e4 		 	& 358.5$^5$		& 63.3	 		& 68.3		 & 98.7      & 120.4 			&			&					\\
Canes Ven. I  & 2.30e5 		 	& 74.3 			& 79.8	 		& 218  		 & 254.1     & 78	 			&			&					\\
Carina		  & 3.80e5 			& 264.8$^5$		& -22.5	 		& 104  		 & 60.8      & -13.8  	 		&			& this work		\\
Car2 		  & 1.08e4$^\star$ 	& 270.0$^1$ 	& -17.1  		& 36.2		 & 25.3      & 211.4$^\dagger$  &			& K18				\\
Car3		  & 1.56e3$^\star$ 	& 270.0$^1$ 	& -16.8  		& 27.8		 & 62.9      & 51.6$^\dagger$   &			& K18				\\
Cra2		  & 3.26e5$^\star$ 	& 283.8$^2$ 	& 42.0 	 		& 116.9		 & 114.5     & -79.1$^\dagger$	&			&					\\
Draco		  & 2.90e5 		 	& 79.2$^5$		& 34.3	 		& 77.2		 & 125.8     & -74.6			&			&					\\
Dra2		  & 2.47e3$^\star$ 	& 98.3$^1$ 		& 42.9	 		& 22.3$^2$   & 125.4     & -159$^\dagger$  	& K18		&					\\
Eri3		  & 1.08e3$^3$ 		& 274.3$^2$ 	& -59.6  		& 87.1		 & 48.2      & 		 			& S17		&					\\
Fornax		  & 2.00e7 		 	& 243.84$^5$    & -67.1  		& 146.1      & 114.5     & -38.2			&			&  this work		\\
Hercules	  & 3.70e4 		 	& 28.7 			& 36.9	 		& 126  		 & 158.5     & 145	 			&			&					\\
Hor1	 	  & 3.92e3$^3$		& 270.9$^2$ 	& -54.9  		& 79.3		 & 38.5      & -30.4$^\dagger$	&			& K18, S17			\\
Hor2		  & 1.88e3$^\star$ 	& 262.5$^2$ 	& -54.1  		& 79		 & 39.6      & 		 			& S17		&					\\
Hyd1	 	  & 1.30e4$^\star$	& 304.5$^1$ 	& -16.5  		& 27.6		 & 28.4      & -51.4$^\dagger$	&			& K18				\\
Hyd2		  & 1.42e4$^3$ 	 	& 295.6$^2$ 	& 30.5	 		& 125.2		 & 113.5     & 134.2$^\dagger$ 	& K18		&					\\
Leo I		  & 5.50e6 		 	& 228.1$^5$		& 50.1	 		& 250.5		 & 263.7     & 159.6 			&			&		\\
Leo II		& 7.40e5 		 	& 223.7$^5$ 	& 68.6	 		& 231.2	     & 255.2     & 18.6	 			&			&			\\
LMC	        & 1.50e9		 	& 290.2$^5$		& -32.5	 		& 51.8	     & 0.0	     & 43.4	 			&			&			\\
Phx2		& 2.25e3$^3$	 	& 323.3$^2$ 	& -60.2  		& 80.2	     & 51.9	     & 		 			& K18, S17	&			\\
Ret3$^2$  	& 2.00e3 		 	& 273.9  		& -45.7  		& 92	     & 45.0	     & 		 			& S17		&			\\
Sag2		  & 2.06e4$^\star$ 	& 18.9$^2$ 		& -22.9  		& 60.1 	     & 49.6	     & 		 			&			&			\\
Sculptor	& 2.30e6 		 	& 318.6$^5$		& -80.2	 		& 86.7	     & 65.6	     & 77.0	 			&			& 			\\
Sextans I	& 4.40e5 		 	& 250.3 		& 43.7	 		& 83.8	     & 94.1	     & 37.7  			&			&			\\
SMC			& 4.60e8 			& 309.4$^5$		& -41.8	 		& 66.2	     & 23.2	     & -4.2 	 		&			& K18, S17	\\
Tuc4$^2$  	& 2.20e3 		 	& 313.3  		& -55.3  		& 45.5	     & 26.6	     & 		 			& S17		&			\\
Tuc5$^2$  	& 5.00e2 		 	& 316.3  		& -51.9  		& 51.9	     & 28.2	     & 		 			& S17		&			\\
Ursa Minor	  & 2.90e5 		 	& 96.5$^5$		& 45.6	 		& 75	     & 125.4   	 & -72.4	 		&			&			\\

\end{tabular*}

\caption{Properties of dwarf galaxies near the Milky Way, selected as having either \mstr > 3\e{4} \msun, or begin determined a potential LMC satellite by S17, K18, or this work. All units and coordinates are in the galactocentric frame. The columns `possible' and `confirmed' refer to the LMC association criteria established by \citet[`S17']{sales2017truMCsats} and \citet[`K18']{kallivayalil2018MCsats}. Galaxies determined to be consistent with the LMC system by our calculations of angular momenta (listed in Table \ref{jtable}) from Gaia DR2 data \citep{helmi2018gaia} are listed as `this work.' Any \mstrs with a star marker ($^\star$) was calculated from the visual magnitude listed in K18 (except for Antlia 2, whose $M_V$ was obtained from \citealt{torrealba2018antlia2}) assuming a mass-to-light ratio of 2. Any $V_r$ with a dagger ($^\dagger$) was converted from its originally tabulated heliocentric value in K18. Numbered superscripts refer to sources. Any row with no superscripts is from \citealt{mcconn2012DG}. A superscript on the name of the galaxy means all properties came from that source. If a property has no superscript but there are others in its row, the nearest superscript to the left is its source. References: 1 - K18; 2 - S17; 3 - \citealt{ferrero2012dwarfhalos}; 4 - \citealt{koposov2018hyi}; 5 - \citealt{helmi2018gaia}; 6 - \citealt{torrealba2018antlia2} }
\label{tab:obs}
\end{table*}

\bsp	
\label{lastpage}
\end{document}